\documentstyle[aps,preprint,pra,url]{revtex}
\begin{document}
\newcommand{\etal}{{\em et al.}\/}
\newcommand{\IP}{inner polarization}
\newcommand{\IPF}{\IP\ function}
\newcommand{\IPFs}{\IP\ functions}
\newcommand{\auth}[2]{#1 #2, }
\newcommand{\jcite}[4]{#1 {\bf #2}, #3 (#4)}
\newcommand{\et}{ and }
\newcommand{\twoauth}[4]{#1 #2 and #3 #4,}
\newcommand{\oneauth}[2]{#1 #2,}
\newcommand{\andauth}[2]{and #1 #2, }
\newcommand{\book}[4]{{\it #1} (#2, #3, #4)}
\newcommand{\erratum}[3]{\jcite{erratum}{#1}{#2}{#3}}
\newcommand{\inpress}[1]{{\it #1}}
\newcommand{\inbook}[5]{In {\it #1}; #2; #3: #4, #5}
\newcommand{\JCP}[3]{\jcite{J. Chem. Phys.}{#1}{#2}{#3}}
\newcommand{\jms}[3]{\jcite{J. Mol. Spectrosc.}{#1}{#2}{#3}}
\newcommand{\jmsp}[3]{\jcite{J. Mol. Spectrosc.}{#1}{#2}{#3}}
\newcommand{\jmstr}[3]{\jcite{J. Mol. Struct.}{#1}{#2}{#3}}
\newcommand{\cpl}[3]{\jcite{Chem. Phys. Lett.}{#1}{#2}{#3}}
\newcommand{\cp}[3]{\jcite{Chem. Phys.}{#1}{#2}{#3}}
\newcommand{\pr}[3]{\jcite{Phys. Rev.}{#1}{#2}{#3}}
\newcommand{\jpc}[3]{\jcite{J. Phys. Chem.}{#1}{#2}{#3}}
\newcommand{\jpcA}[3]{\jcite{J. Phys. Chem. A}{#1}{#2}{#3}}
\newcommand{\jpca}[3]{\jcite{J. Phys. Chem. A}{#1}{#2}{#3}}
\newcommand{\jpcB}[3]{\jcite{J. Phys. Chem. B}{#1}{#2}{#3}}
\newcommand{\jpB}[3]{\jcite{J. Phys. B}{#1}{#2}{#3}}
\newcommand{\PRA}[3]{\jcite{Phys. Rev. A}{#1}{#2}{#3}}
\newcommand{\PRB}[3]{\jcite{Phys. Rev. B}{#1}{#2}{#3}}
\newcommand{\PRL}[3]{\jcite{Phys. Rev. Lett.}{#1}{#2}{#3}}
\newcommand{\jcc}[3]{\jcite{J. Comput. Chem.}{#1}{#2}{#3}}
\newcommand{\molphys}[3]{\jcite{Mol. Phys.}{#1}{#2}{#3}}
\newcommand{\mph}[3]{\jcite{Mol. Phys.}{#1}{#2}{#3}}
\newcommand{\APJ}[3]{\jcite{Astrophys. J.}{#1}{#2}{#3}}
\newcommand{\cpc}[3]{\jcite{Comput. Phys. Commun.}{#1}{#2}{#3}}
\newcommand{\jcsfii}[3]{\jcite{J. Chem. Soc. Faraday Trans. II}{#1}{#2}{#3}}
\newcommand{\FD}[3]{\jcite{Faraday Discuss.}{#1}{#2}{#3}}
\newcommand{\prsa}[3]{\jcite{Proc. Royal Soc. A}{#1}{#2}{#3}}
\newcommand{\jacs}[3]{\jcite{J. Am. Chem. Soc.}{#1}{#2}{#3}}
\newcommand{\joptsa}[3]{\jcite{J. Opt. Soc. Am.}{#1}{#2}{#3}}
\newcommand{\cjc}[3]{\jcite{Can. J. Chem.}{#1}{#2}{#3}}
\newcommand{\ijqcs}[3]{\jcite{Int. J. Quantum Chem. Symp.}{#1}{#2}{#3}}
\newcommand{\ijqc}[3]{\jcite{Int. J. Quantum Chem.}{#1}{#2}{#3}}
\newcommand{\spa}[3]{\jcite{Spectrochim. Acta A}{#1}{#2}{#3}}
\newcommand{\tca}[3]{\jcite{Theor. Chem. Acc.}{#1}{#2}{#3}}
\newcommand{\tcaold}[3]{\jcite{Theor. Chim. Acta}{#1}{#2}{#3}}
\newcommand{\jpcrd}[3]{\jcite{J. Phys. Chem. Ref. Data}{#1}{#2}{#3}}
\newcommand{\science}[3]{\jcite{Science}{#1}{#2}{#3}}
\newcommand{\CR}[3]{\jcite{Chem. Rev.}{#1}{#2}{#3}}
\newcommand{\bbpc}[3]{\jcite{Ber. Bunsenges. Phys. Chem.}{#1}{#2}{#3}}
\newcommand{\acie}[3]{\jcite{Angew. Chem. Int. Ed.}{#1}{#2}{#3}}
\newcommand{\ijck}[3]{\jcite{Int. J. Chem. Kinet.}{#1}{#2}{#3}}
\newcommand{\jct}[3]{\jcite{J. Chem. Thermodyn.}{#1}{#2}{#3}}

\newcommand{\deltah}[0]{$\Delta$H}
\newcommand{\deltahf}[0]{$\Delta$H$_f$}
\newcommand{\hoof}[0]{$\Delta$H$_f$$^{298}$}
\newcommand{\hof}[0]{$\Delta H^\circ_f$}
\newcommand{\hofzero}[0]{$\Delta$H$_f$$^{0}$}
\newcommand{\abin}{{\em ab initio}}
\newcommand{\tabscfextrapref}{I}
\newcommand{\tabeageetwooneref}{II}
\newcommand{\tabeadgeetwooneref}{III}
\newcommand{\tabsoref}{IV}
\newcommand{\tabeageetwotworef}{S-I}
\newcommand{\tabeadgeetwotworef}{V}
\newcommand{\tabipgeetwooneref}{VI}
\newcommand{\tabipdgeetwooneref}{VII}
\newcommand{\tabmethaneplusfreqref}{VIII}
\newcommand{\tabipgeetwotworef}{S-II}
\newcommand{\tabipdgeetwotworef}{IX}
\newcommand{\tabhofgeetwooneref}{S-III}
\newcommand{\tabhofdgeetwooneref}{X}
\newcommand{\tabhofgeetwotworef}{S-IV}
\newcommand{\tabhofdgeetwotworef}{XI}
\newcommand{\tabPAdecompref}{S-V}
\newcommand{\tabPAtableref}{XII}

\draft
\title{Assessment of W1 and W2 theories for the computation of
electron affinities, ionization potentials, heats of formation,
and proton affinities\thanks{Dedicated to the memory of Professor Shneior Lifson z$"$l (March 18, 1914--January 22, 2001)}}
\author{Srinivasan Parthiban and Jan M.L. Martin\thanks{Corresponding author. {\rm E-mail:} {\tt comartin@wicc.weizmann.ac.il}}}
\address{Department of Organic Chemistry,
Kimmelman Building, Room 262,
Weizmann Institute of Science,
IL-76100 Re\d{h}ovot, Israel
}
\date{{\em J. Chem. Phys.} MS {\bf 303115JCP}; Received Oct. 31, 2000; Accepted Jan. 24, 2001}
\maketitle
\begin{abstract}
The performance of two recent {\em ab initio} computational thermochemistry
schemes, W1 and W2 theory [J.M.L. Martin and G. de Oliveira, \JCP{111}{1843}{1999}], 
is assessed for an enlarged sample of thermochemical data consisting 
of the ionization potentials and electron affinities in the G2-1 and 
G2-2 sets, as well as the heats of formation in the G2-1 and a subset 
of the G2-2 set. We find W1 theory to be several times more accurate
for ionization potentials and electron affinities than commonly used
(and less expensive) computational thermochemistry schemes such as 
G2, G3, and CBS-QB3: W2 theory represents a slight improvement for 
electron affinities but no significant one for ionization potentials.
The use of a two-point $A+B/L^5$ rather than a three-point $A+B/C^L$
extrapolation for the SCF component greatly enhances the numerical
stability of the W1 method for systems with slow basis set convergence.
Inclusion of first-order spin-orbit coupling is essential for accurate
ionization potentials and electron affinities involving degenerate 
electronic states: inner-shell correlation is somewhat more important
for ionization potentials than for electron affinities, while scalar
relativistic effects are required for the highest accuracy. The 
mean deviation from experiment for the G2-1 heats of formation is 
within the average experimental uncertainty. W1 theory appears to
be  a valuable tool for obtaining benchmark quality proton affinities.
\end{abstract}
\newpage
\section{Introduction}
Development of models based on molecular orbital theory for theoretical 
thermochemistry involves five key steps\cite{pople99}: 
defining a target accuracy, formulation of theory, implementation through 
programs, validating the models against reliable experimental values and 
prediction on any molecular system by the end user. 
At present, only {\em ab initio} methods can claim "chemical 
accuracy" (commonly defined as 1 kcal/mol) for small and medium 
sized molecules. The most popular such methods are the Gaussian-$n$  (G$n$)
theories\cite{g1paper,g2paper,g3paper} of Pople and coworkers (which are based 
on a combination of additivity approximations and empirical corrections 
applied to relatively low-level calculations), followed by the CBS 
approaches\cite{cbs1,cbs2,cbs3} of Petersson and coworkers which are 
intricate combinations of extrapolation and empirical correction 
schemes. 

Very recently, Martin and de Oliveira presented two theoretical
thermochemistry schemes known as W1 and W2 (Weizmann-1 and Weizmann-2)
theory\cite{w1}, which aim at `benchmark accuracy', defined by these
authors as 1 kJ/mol (0.24 kcal/mol). For a set of 28 experimentally
very precisely known molecular total atomization energies, the
more cost-effective of the two schemes, W1 theory, achieved a mean
absolute error of 0.37 kcal/mol, while the more rigorous of the
two schemes, W2 theory, achieved a mean absolute error of 0.23 
kcal/mol. (It should be pointed out that these methods are free
of parameters derived from experiment: W1 theory does contain one 
parameter --- the exponent for the valence correlation 
extrapolation --- that
is derived from W2 calculations, not experiment.)
Martin later proposed a minor modification of W1 theory,
denoted W1$'$ theory\cite{so3tae}, which appeared to yield considerably improved results for
second-row compounds at no additional cost. 
(For first-row compounds, it is identical to W1 theory.)

In recent years, density functional theory (DFT) methods have also
reached a level of sophistication where they can provide thermochemical
data to within a few kcal/mol, notably the hybrid B3LYP (Becke 3-parameter
exchange with Lee-Yang-Parr correlation\cite{Bec93,LYP}) and 
B97 (Becke 1997\cite{Bec97}) exchange-correlation functionals, but
also the `pure DFT' HCTH (Hamprecht-Cohen-Tozer-Handy\cite{hcth}) 
exchange-correlation functional. A recent collection of reviews on
computational thermochemistry methods may be found in an ACS
Symposia volume edited by Irikura and Frurip.\cite{acs677}

The most fundamental thermochemical property of a compound, from an 
experimental point of view, is its heat of formation (\hof\ ) in the 
gas phase. From a computational chemistry point of view, the total atomization 
energy (TAE, $\Sigma D_0$)\cite{martin-en} is the most fundamental such 
quantity. Using the experimental heats of formation of the atoms in 
the gas phase, TAEs can be directly related to the gas-phase heats of formation. 

Prior to proper application of any new model by the end user, it should
be tested against known high quality experimental results. For this purpose,
Pople and coworkers proposed two standard test sets of thermochemical data:
the G2-1 test set\cite{g2paper} being the smaller and containing small molecules,
and the G2-2 test set\cite{g2hof,g2ipea} containing larger systems. 
These sets of thermochemical data, covering 148 neutral and 146 ionic species,
have been used fairly extensively (e.g.\cite{Fel99}) 
to test the performance of various
computational thermochemistry methods, notably the G$n$ theories and their
variants\cite{g2hof,g2ipea,aleph,beit,gimel,dalet}, 
density functional methods\cite{g2hof,g2ipea,charlie95,deproft97},
and the CBS family of methods\cite{cbs1,cbs2,cbs3,cbs4}. 

The main problem with the G2-1 and G2-2 test sets for heats of formation
is the limited accuracy
of the experimental data themselves. These were critically reviewed by
Liebman and Johnson\cite{liebman}, who concluded that less than half of
the data even met the less rigorous 1 kcal/mol accuracy criterion. For
methods of the W1/W1$'$/W2 type, this is clearly a major impediment to
their validation for a larger experimental data set, and alternatives
need to be sought.

Pople and coworkers also defined G2-1\cite{g2paper} and G2-2\cite{g2ipea} 
data sets for ionization
potentials and electron affinities. The accuracy of these experimental
data is much more satisfactory, and it could be argued that they are
in fact more suitable test sets  for the 
validation of theoretical thermochemistry methods. While ionization
potentials are comparatively easy to reproduce well, electron
affinities are a very taxing test for any electronic structure
method. This is true
both in terms of the basis set (addition of the electron
entails a profound change in the spatial extent of the wave function)
and in terms of the electron correlation method (effectively, the
number of interacting electrons in the system changes).

The purpose of the present paper is to assess the performance of 
W1 and W2 theory for an extended data set of thermochemical data.
Specifically, for W1 theory we shall consider the G2-1 and G2-2
datasets for ionization potentials and electron affinities,
as well as the G2-1 and a subset of the G2-2 data set for 
heats of formation. W2 theory will be considered for the G2-1
data set for IPs, EAs, and heats of formation.
Finally we shall turn to W1 and W2 theory for proton affinities.

\section{Computational methods}

Geometry optimizations and vibrational frequency
calculations using the B3LYP (Becke 3-parameter-Lee-Yang-Parr\cite{Bec93,LYP})
density functional method have been carried out using Gaussian 98 revision A7\cite{g98revA7}. 
(Following the recommendations in Ref.\cite{grids} larger grids than the default 
were used in the DFT calculations if necessary, specifically a pruned (99,590) grid 
for integration and gradients, and a pruned (50,194) grid for the solution of the 
coupled perturbed Kohn-Sham equations.)
All
other calculations were carried out using MOLPRO 98.1\cite{molpro98}
and a driver for the W1/W2 calculations\cite{autoW1W2}
written in MOLPRO's scripting language. The lion's share of the
calculations was carried out on a Compaq ES40 with four 667 MHz 
Alpha EV67 CPUs, and a scratch volume consisting of six 18GB SCSI 2 ultrawide
disks striped in software. Remaining calculations were carried out on the
SGI Origin 2000  of the Faculty of Chemistry.

The SCF and valence correlation basis sets are Dunning's 
augmented correlation consistent $n$-tuple zeta\cite{Dun89,Ken92,Dun97}
(aug-cc-pV$n$Z) basis sets; for second-row atoms, 
high-exponent $d$
and $f$ functions were added (denoted '+2d' or '+2d1f')
as was found repeatedly\cite{Bau95,so2} to be necessary for accommodating
inner-shell correlation effects. (Unless indicated otherwise,
regular `un-augmented' cc-pV$n$Z basis sets are used for H, Li, and Be,
as well as cc-pV$n$Z+2d1f for Na and Mg.)
For all remaining steps (inner-shell
correlation, scalar relativistic effects, and spin-orbit coupling),
the MTsmall (Martin-Taylor small\cite{w1}) core correlation basis set
was used. Restricted open-shell wave functions were used throughout for
open-shell species.

The W1/W2 energy consists of seven components, each of which we shall
detail in turn for reasons of clarity and self-containedness. 

\noindent {\bf (0)} Reference geometries are obtained at the 
B3LYP/cc-pVTZ+1 level in the case of W1 theory, and at the 
CCSD(T)/cc-pVQZ+1 level in the case of W2 theory.
In both cases, the `+1' signifies the addition\cite{sio}
of a high-exponent
$d$ function to second-row elements, the exponent having been set equal
to the highest $d$ exponent in the corresponding cc-pV5Z basis set.

\noindent {\bf (1)} In the original W1 and W2 papers, the SCF limit
was obtained by geometric extrapolation\cite{Fel92}
\begin{equation}
TAE_{\rm SCF}(n)=TAE_{SCF,\infty}+A.B^n
\end{equation}
of the molecular total atomization energy TAE computed 
using cc-pV$n$Z+2d1f basis sets, where for W2 theory $n=\{T,Q,5\}$
(with $l$=\{3,4,5\}), and for W1 theory $n=\{D,T,Q\}$
(with $l$=\{2,3,4\}). (In practice, this means that
$TAE_{SCF,\infty}=TAE_{SCF,n}-(TAE_{SCF,n}-TAE_{SCF,n-1})^2/(TAE_{SCF,n}-2TAE_{SCF,n-1}+TAE_{SCF,n-2})$.
The exponential convergence behavior of the
SCF energy has repeatedly been demonstrated empirically
(e.g. by Jensen\cite{JensenSCF} and by Martin and Taylor\cite{n2h2})
in comparisons with numerical Hartree-Fock energies for small
molecules.

The geometric formula, for this particular application, has the
minor disadvantage that its extrapolated limit depends on whether
the extrapolation is carried out on TAE, or on the constituent
energies. (In practice the differences are quite minor.) Based on 
the asymptotic convergence behavior\cite{Sch63,Hil85,Kut92}
of the pair energy in an electron
pair that does not have an interelectronic cusp, 
Petersson and coworkers\cite{PetSCF} previously considered 
(within the context of their CBS family of methods)
an alternative formula 
$E_{\rm SCF}(n)=E_{SCF,\infty}+\sum_{l=n+1}^{\infty}\frac{A}{l+1/2}^6$.
(Using Euler-Maclaurin summation, we find this to be equivalent
to the simple two-point 
formula $E_{\rm SCF}(n)=E_{SCF,\infty}+A/n^5+O(n^{-7})$.)
Martin and Taylor\cite{c2h4tae}
previously considered the difference between the
three-point geometric formula (with $n$=\{T,Q,5\}) and the two-point
formula (with $n$=\{Q,5\}) for a small set of molecules
and found the differences to be negligible.
We also find this to be the case for the much larger sample of molecules
surveyed here.

However, when considering three-point  geometric $n$=\{D,T,Q\} versus
two-point $n$=\{T,Q\} in the present work, we found that, while the
differences are quite small for almost all first-row and most second-row
systems, significant differences (in excess of 1 kcal/mol) exist for
a few first-row systems (e.g. LiF) 
and a rather larger number of second-row systems (e.g. many silicon
compounds). Some conspicuous examples can be found in Table \tabscfextrapref.
Upon closer inspection, this was revealed to be caused by
the three points lying nearly on a straight line, causing the
geometric extrapolation to yield erratic results. The two-point
extrapolation is invariably closer to the extrapolated limit
obtained from the larger basis sets: in unproblematic cases, it
yields essentially the same results as the three-point extrapolation.
As a result, we are recommending that the two-point $A+B/l^5$ extrapolation
be used in W1 and W2 theory from now on: one beneficial side effect is
that the extrapolated limit for this two-point formula is easily seen 
to be independent of whether the extrapolation is carried out on the
molecule or the constituent atoms. Since the SCF component was the only
component for which such an ambiguity existed in the 
original W1 and W2 theory, this permits the quoting of "total W1 and
W2 energies" for arbitrary systems.

It is also seen in Table \tabscfextrapref\ that the main argument in favor
of W1$'$ theory (in which the
AVTZ+2d1f basis set is replaced by an AVTZ+2d basis set, for balance 
reasons\cite{so3tae}) over standard W1 theory, namely 
an SCF limit in better
agreement with that obtained from larger basis sets, appears to be
obviated by the new extrapolation. We
shall not consider W1$'$ theory further in the course of this paper.

\noindent {\bf (2)} In the W2 case, the CCSD 
(coupled cluster with all singles and
doubles\cite{Pur82}) valence correlation contribution to TAE
is obtained using the aug-cc-pVQZ+2d1f and aug-cc-pV5Z+2d1f
basis sets, then extrapolated to the infinite basis limit using
the expression\cite{Hal98} $E(l)=E_\infty+A/L^3$. (In practice, this 
means $E_\infty=E_l+(E_l-E_{l-1})/((\frac{l}{l-1})^3-1)$.) The
arguments in favor of this expression (derived from the 
known asymptotic convergence behavior of the interelectronic
cusp\cite{Sch63,Hil85,Kut92}) have been detailed at length
elsewhere\cite{w1,Hal98,Bak2000} and will not be repeated here.
In the W1 case, the unmodified expression leads to systematically
overestimated correlation contributions to TAE\cite{w1}: here we employ
$E_\infty=E_l+(E_l-E_{l-1})/((\frac{l}{l-1})^\alpha-1)$, where
$\alpha=3.22$ was determined\cite{w1} to yield the best agreement
with the extrapolated W2 CCSD energies.
Both for W1 and W2 theory, the largest basis set CCSD calculation
is carried out (except for very small systems)
using the direct CCSD algorithm of Sch\"utz, Lindh, and Werner\cite{dirccsd}
as implemented in MOLPRO98.1.

\noindent {\bf (3)} The contribution of connected triple excitations is 
obtained at the CCSD(T) level (CCSD with a quasiperturbative a posteriori
correction for connected triple excitations\cite{Rag89}). 
As the $T_3$ contribution
is known\cite{Hel97}
to converge more rapidly than the contribution of $\exp(T_1+T_2)$, hence
this contribution is obtained from CCSD(T) calculations with the smaller
two basis sets and extrapolated to the infinite-basis limit using 
$E_\infty=E_l+(E_l-E_{l-1})/((\frac{l}{l-1})^\alpha-1)$, where, as
for the CCSD energy, $\alpha$=3 exactly for W2 theory and $\alpha$=3.22
for W1 theory. (For open-shell systems, the definitions of the restricted
CCSD and CCSD(T) energy as given in Ref.\cite{Wat93} has been used.)

\noindent {\bf (4)} The inner-shell correlation contribution is computed as
the difference between CCSD(T)/MTsmall\cite{w1} values with and without
constraining the inner-shell orbitals to be doubly occupied. (In the
case of the second-row elements, the very deep-lying
(1s)-like orbitals are constrained 
to be doubly occupied throughout.)

\noindent {\bf (5)} The scalar relativistic contribution
is computed as expectation values of the one-electron Darwin and mass-velocity (DMV)
operators\cite{Cow76,Mar83} for the ACPF/MTsmall (averaged coupled
pair functional\cite{Gda88}) wave function, with all inner-shell
electrons correlated except the (1s)-like orbitals of second-row elements.
Bauschlicher\cite{Bau2000} demonstrated that, for first- and second-row
systems, this approach yields essentially
identical results to more rigorous relativistic calculations.

\noindent {\bf (6)} For closed-shell systems, or open-shell systems in nondegenerate
electronic states, there is no molecular first-order spin-orbit contribution,
and the contribution to TAE is merely the sum of the atomic spin-orbit
corrections. For open-shell systems in degenerate states, we have calculated
spin-orbit corrections at the CISD level with the MTsmall basis set, and 
again correlating all electrons except for the (1s) on second-row elements.

\noindent {\bf (7)} The molecular zero-point energy and thermal corrections
were obtained at the B3LYP/cc-pVTZ+1 level. The zero-point energies within
the harmonic approximation are scaled by 0.985, primarily to correct
for anharmonicity. The scale factor was obtained\cite{w1} by comparison
with experimental (or high-level theoretical) anharmonic zero-point energies
for 28 molecules.

(Adiabatic) electron affinities (EAs) are calculated as the difference between the 
TAE$_0$ values of the anion and the corresponding
neutral species, at their respective optimized geometries:
\begin{equation}
EA_0 = TAE_0({\rm anion}) - TAE_0({\rm neutral})
\end{equation}
Likewise, the (adiabatic) ionization potentials (IPs)
are calculated as the difference
in total atomization energies at 0 K of the cation and the corresponding
neutral, at their respective optimized geometries:
\begin{equation}
IP_0 = TAE_0({\rm neutral}) - TAE_0({\rm cation})
\end{equation}

Theoretical heats of formation at 0 K were calculated by subtracting the W$n$
calculated TAE$_0$ ($\Sigma D_0$) value from experimental enthalpies
of formation of the isolated atoms. For any molecule, such as A$_x$B$_y$H$_z$, the
heat of formation at 0 K is given by
\begin{equation}
 \Delta H_f^{0} (A_xB_yH_z,0 K) = x\Delta H_f^{0} (A, 0 K) + y\Delta H_f^{0} (B, 0 K) +
 z\Delta H_f^{0} (H, 0 K) - \Sigma D_0.
\end{equation}
The CODATA\cite{Cod89} values of the atomic $\Delta H_f^{0}$ are used with the
exception of boron and silicon, for which we have used revised
values recommended by Bauschlicher, Martin, and Taylor\cite{bf3,bf3cwb} for boron and by
Martin and Taylor\cite{martin-si99} for silicon. Theoretical
heats of formation at 298 K (\hof) are calculated by
correction to $\Delta H_f^{0}$ as follows:

\begin{eqnarray}
\Delta H_f^{0} (A_xB_yH_z, 298 K)&=& \Delta H_f^{0} (A_xB_yH_z,0 K) \nonumber\\
                &+& [H^o (A_xB_yH_z, 298 K) - H^o (A_xB_yH_z, 0 K)] \nonumber\\
                &-& x {\rm hcf}_{298}[{\rm A, st}] - y {\rm hcf}_{298}[{\rm B, st}] 
                - z {\rm hcf}_{298}[{\rm H, st}] 
\end{eqnarray}
While the enthalpy functions $hcf_{T}\equiv H_{T}-H_0$ for the molecule are obtained
using the RRHO (rigid rotor-harmonic oscillator) approximation from
the unscaled B3LYP/cc-pVTZ+1 vibrational frequencies, the enthalpy
functions for the standard states of the elements are taken directly from CODATA.

Proton affinities (PA) are obtained from the total atomization energies
at 0 K as follows:
\begin{equation}
PA_0(B)  = TAE_0({\rm BH^+}) - TAE_0({\rm B})
\end{equation}
Finally, PAs at 298 K are calculated by correction to PA$_0$ as follows:
\begin{equation}
PA_{298}(B) = PA_0(B) + {\rm hcf}_{298}(BH^+) - {\rm hcf}_{298}(B) - \frac{5RT}{2}
\end{equation}
where $5RT/2$ is the enthalpy function of the $H^+$ ion.

\section{Test sets used}

The original G2-1 ion test set consists of 25 EAs and 38 IPs while the G2-2 
test set included 33 EAs and 50 IPs. In the G3 paper, Curtiss 
\etal\cite{g3paper} applied G3 theory to the G2-1 and G2-2 test sets, minus 
three ionization potentials (due to the size of the molecules concerned). 
In this study we exclude five additional ionization potentials and one electron 
affinity from the G2-2 test set, for the same reason.
Both W1 and W2 theories were evaluated for the G2-1 test set, while 
only W1 theory was considered for the G2-2 test set.
For the purpose of evaluation of \hof, 
the original G2-1 neutral test set consists of 55 molecules while the 
G2-2 test set includes 93 molecules. Again, we have considered both 
W1 and W2 theories for the G2-1 neutral test set of molecules and compared the results 
with G2, G3 and CBS-Q values. (For the W2 calculations, three species were omitted
because of their size.) It should be noted that the G2-2 test set 
contains several fairly large molecules and some of the experimental 
\hof\ for the species in G2-2 test set possess large uncertainties as well as several
experimental values spanning a wide range. Therefore, we have selected a subset of 
27 out of the 93 G2-2 neutral molecules, which are tractably small and 
for which the experimental enthalpies are reasonably accurate. To these molecules
we applied W1 theory, and to a subset of them W2 theory.

\section{Results and Discussion}
\subsection{Electron Affinities}

We shall first consider the G2-1 test set. A breakdown of the
different components of the W1 values is given in Table \tabeageetwooneref,
while a comparison between various levels of theory 
(including W1, W2, G2, G3, and CBS-Q) and experiment
is given in Table \tabeadgeetwooneref.

We note first that a substantial number of species have negative 
electron affinities at the SCF level: the binding of the electron 
results from the additional correlation energy in those cases. 
Inclusion of connected triple excitations is essential. In contrast,
inner-shell correlation does not appear to be very important for the
G2-1 EAs. Scalar relativistic effects are somewhat more important: 
with the exception of CH$_{2}$, CH$_{3}$, and SiH$_{3}$ they 
uniformly decrease the electron affinity. As expected, the scalar 
relativistic effect is somewhat more important in second-row than in 
first-row systems. The change in the zero-point energies can be
fairly substantial, particularly for hydrides.

For heats of formation of closed-shell systems (and open-shell 
systems with nondegenerate ground states), the molecular first-order 
spin-orbit splitting vanishes, reducing the spin-orbit correction to
the sum of the corrections for the constituent atoms. Since a fair
number of the species in the G2-1 set have a degenerate state
for either the neutral or the anionic system (or in fact for both),
some account for molecular spin-orbit coupling cannot be avoided. 
We have considered a number of (relatively) inexpensive approximations
within the MTsmall basis set used for the core correlation and scalar
relativistic steps, including SCF, CISD, and CISD with inner-shell
electrons correlated. The computed corrections at these levels of 
theory for a number of (neutral, cationic, and anionic) species have
been compared in Table \tabsoref\ with values obtained from experimental 
fine structures. For most first-row species, satisfactory results are
already obtained at the SCF and definitely at the CISD level; for the 
second-row species, correlation from the (2s2p) inner-shell orbitals
appears to be essential, as was previously found by de Oliveira et 
al.\cite{glen99} for the 2nd-row atoms 
and by Nicklass et al.\cite{Nic2000} in a convergence study for the
halogen atoms. (The rather
weak basis set dependence found by these latter authors\cite{Nic2000}
is consistent with our own findings.) Only for the ClO molecule we find
a substantial error: inspection of the spectroscopic
constants\cite{Hub79} for the few lowest states reveals that the 
$X~^2\Pi$ ground state in fact has a anomalously small splitting
compared to the $A~^2\Pi$ state; since this latter state mixes in quite 
prominently into the $X~^2\Pi$ wave function, the splitting is 
severely underestimated unless the $A~^2\Pi$ state is admitted to the 
zero-order wave function. Using a CASSCF reference space consisting 
of the valence orbitals except for the Cl(3s) and O(2s) like orbitals,
and supplemented with the first Rydberg $\pi$ orbitals, yields a 
spin-orbit
correction in excellent agreement with experiment.

As can be seen from Table \tabeageetwooneref, these spin-orbit corrections 
are in fact essential for good agreement with experiment for several 
of the systems. At the W1 level, we find a mean absolute discrepancy
(MAD) from experiment of 0.016 eV (Table \tabeadgeetwooneref), which is a quite substantial
improvement over the G2 (0.057 eV), G3 (0.049 eV), and CBS-QB3 (0.054 
eV) values. Perhaps even more importantly, the maximum error is
likewise much smaller, 0.051 eV for CH$_{3}$, followed by 0.043 eV for SiH$_{2}$.
In the case of CH$_{3}$, not only is the electron affinity very small
(G3 and CBS-QB3 in fact predict the wrong sign), but the 
harmonic 
approximation for the zero-point energy is of dubious reliability
(see Schwenke\cite{Schwenke99CH3}).
W2 represents a minor improvement over W1, at vastly greater 
computational expense: MAD=0.012 eV. Using even larger basis sets,
de Oliveira et al.\cite{glen99} found the mean absolute error for the
atoms H, B--F, and Al--Cl to be 0.009 eV at the CCSD(T) level; by 
employing CCSDT and full CI corrections, this error could be reduced by
an order of magnitude. (The importance of these corrections was about
evenly split between higher-order $T_3$ effects and effects of 
connected
quadruple excitations, $T_{4}$.)
We conclude that the accuracy of W2 theory
(and, to a lesser extent, W1 theory) is mostly determined by the 
imperfections in the CCSD(T) method.

We shall now consider the G2-2 set of electron affinities, for which
only W1 (not W2) calculations were carried out. A comparison with
other theoretical thermochemistries and with experiment is given in
Table \tabeadgeetwotworef, while a breakdown of
contributions is given in Table \tabeageetwotworef\cite{epaps}.

The trends seen for the G2-1 set largely continue for the G2-2 set. 
However, inner-shell correlation is somewhat more important for some
species (e.g. Al, because of the small subvalence/valence gap, C$_{2}$,
and S$_{2}$O). One exception to the general trends is that electron
correlation in fact {\em decreases} EA(C$_{2}$): this is an artifact of
the multireference character of the $X~^1\Sigma^{+}_{g}$ state.
At first sight, scalar relativistic effects seem less important, but
this is an artifact of the relative preponderance of first-row
species compared to the G2-2 set.

Standard W1 results for Li and Na would not involve diffuse functions on 
these low-electronegativity atoms. Not surprisingly, very poor 
electron affinities are thus obtained. We have optimized diffuse 
functions (available in the supplementary material)
for Li, Be, Na, and Mg to accompany the standard cc-pV$n$Z
basis set: the exponents were optimized individually for each angular
momentum at the CISD level for the atomic anion. The W1aug results 
obtained with these basis sets are in near-perfect agreement with 
experiment. (This can reasonably be expected since the electron 
correlation methods used are exact within the finite basis set for the
valence correlation contributions.)

As for the G2-1 set, we find W1 theory to be quite substantially more
reliable than G2 and G3 theory. Substantial discrepancies between W1 
theory and experiment are found for ozone, CH$_{2}$NC, and FO: the 
first two species (and, to a lesser extent, FO) exhibit strong
nondynamical correlation effects, and hence methods that do not 
include corrections for correlation effects beyond CCSD(T) are expected
to yield poor results. G3 theory fortuitously agrees better with 
experiment than W1 theory for these species.

\subsection{Ionization Potentials}

We shall again first consider the G2-1 set. A breakdown of components 
in the W1 computed values can be found in Table \tabipgeetwooneref, while a 
comparison with experiment and less expensive theoretical 
thermochemistry methods can be found in Table \tabipdgeetwooneref.
The relative importance of
correlation is smaller than for the electron affinities: yet in 
absolute terms its contribution is almost as significant as in the EA 
case. While connected triple excitations appear to be somewhat less 
important than for EAs, they can certainly not be neglected with 
impunity. Inner-shell correlation contributions, on the other hand, 
are more important than in the EA case because the valence excitation 
creates a ``hole'' into which core electrons can be excited. 
The large contributions for Na  (4 kcal/mol) and Mg (2 kcal/mol) come 
as no surprise given the small core-valence gap in these atoms. 
Scalar relativistic contributions are important for accurate work: 
with the exception of Li--C and Na--Mg, they consistently lower the IP.
Like for the EAs, we see substantial zero-point effects for the 
hydrides: in the case of CH$_{4}$, this contribution is especially
large because of the known fluxional nature\cite{Davidson1991andrefs} 
of the CH$_{4}^{+}$ cation. And again, spin-orbit splitting is a 
factor to be reckoned with, particularly for such second-row species
as exhibit first-order spin-orbit splitting.

Agreement with experiment is highly satisfactory at the W1 level, 
except for CH$_{4}$ where an atypically large discrepancy is seen. 
Upon inspection, it is revealed that the B3LYP/cc-pVTZ geometry for
CH$_{4}^{+}$ is qualitatively incorrect, exhibiting $D_{2}$ rather
than $C_{2v}$ symmetry. This reflects itself both in an error in the
total energy for the cation and in an error in the zero-point 
contribution. Using a CCSD(T)/cc-pVTZ reference geometry, excellent 
agreement with experiment is in fact obtained.

In an attempt to ascertain whether this issue is specific to the
B3LYP exchange-correlation functional, we carried out geometry
optimizations and vibrational frequency calculations for
CH$_4^+$ using the cc-pVTZ basis set and a variety of exchange-correlation
functionals, including B3P86 (Becke 3-parameter exchange with
Perdew-1986 correlation\cite{Perdew86}), B3PW91 (Becke 3-parameter
exchange with Perdew-Wang-1991 correlation\cite{PW91}), mPW1PW91
(modified Perdew-Wang\cite{Ada98}), mPW1K (Truhlar's empirical
modification of the latter\cite{mPW1K}), BHLYP (Becke 
half-and-half exchange\cite{BHandH} with LYP correlation), BHPW91 (ditto with
PW91 correlation), and BLYP (Becke 1988 exchange\cite{Bec88}
with LYP correlation). Results are summarized in Table \tabmethaneplusfreqref.
Only the functionals with 50\% Hartree-Fock exchange (BHLYP, BHPW91)
or nearly so (mPW1K) find this structure to be a local minimum,
while all other functionals find an imaginary frequency of $a_2$
symmetry. Following the latter downhill leads to the $D_{2d}$ structure.
Given that this behavior persists with a fairly wide variety of correlation
functionals, the problem appears to reside in the exchange functional.
(Note that since W1 theory does not contain any parameters that depend 
on the level of theory for the reference geometry, it can quite well
be carried out from, say, a mPW1K/cc-pVTZ reference geometry for systems which
exhibit this type of problem.)

Mean absolute deviation for W1 is a factor of 3--4 smaller than for 
the inexpensive methods. In this case, only marginal improvement is seen upon going to
the much more expensive W2 method, which is easily 
understood in terms of the faster basis set convergence for the 
cation compared to the anion. Again, we have reason to believe that
the principal factor limiting the accuracy of our calculations are
small deficiencies in the CCSD(T) electron correlation method.

Let us now consider the G2-2 ionization potentials (Tables \tabipdgeetwotworef\ and \tabipgeetwotworef).
Most systems in that set do not exhibit first-order spin-orbit 
splitting, the main exceptions being Ne, Ar, OCS, and 
CS$_{2}$. 
Most trends from the G2-1 set are continued:  one clear
exception to the general rule is CN, for which electron correlation
{\em reduces} the IP. At first sight, this system also exhibits a
large discrepancy of 0.27 eV between theory and experiment, and
discrepancies for the more approximate G2 and G3 methods are similarly high. 
An explanation in terms of the extreme
multireference character of the CN$^{+}$ cation 
would be tempting: however, we repeated the W2 calculation using
full valence CAS-ACPF instead of CCSD(T) at every step, and found
an {\em increase} by 0.04 eV in our computed value (to 13.93 eV). Upon
closer inspection, it appears that the `experimental' IP(CN)=13.6 eV
is in fact a propagated transcription error from Ref.\cite{Ber94}.
The only experimental 
value without a large error bar, 14.03$\pm$0.02 eV, was obtained
by Berkowitz et al.\cite{Ber69} from photoionization data for
HCN$\rightarrow$H$^+$+CN+$e^-$ (19.00$\pm$0.01 eV) and 
HCN$\rightarrow$H+CN$^+$+$e^-$ (19.43$\pm$0.01 eV), as well
as the well-established IP(H). The error bars on the extrapolated
ionization limits may be somewhat optimistic: in addition,
it is well known from benchmark calculations (e.g.\cite{Pet95}: see also
\cite{She98})
that CN$^-$ has a very low-lying $a~^3\Pi$ state ($T_e$=880$\pm$100 cm$^{-1}$\cite{Pet95},
or 0.11 eV). It is not inconceivable that the Berkowitz et al.
value in fact corresponds to generation of the $^3\Pi$ state (especially
since the ground state of HCN$^+$ is $X~^2\Pi$): our W1 and
W2 computed IPs in that case are both 14.07 eV, in excellent
agreement with the Berkowitz value. This problem merits further investigation.

For the purpose of assessment of the error statistics of the 
various methods, however, we have removed IP(CN) from the
sample.  
Significant discrepancies 
--- out of character with the other results --- are then still
seen for 
B$_{2}$H$_{4}$, sec-C$_{3}$H$_{7}$, Si$_2$H$_6$, 
and CH$_{3}$OF. None of these 
species exhibits severe nondynamical correlation, and we note that
there are significant discrepancies between G3 and experiment for all
these species except Si$_2$H$_6$. We suggest that these experimental values may need to be 
reconsidered. In the case of B$_2$H$_4$, it has previously
been suggested\cite{Sta92} that the surface between the $D_{2d}$
structure and the doubly bridged $C_{2v}$ isomer is flat enough that
rigid molecule treatments may not be appropriate. The remaining three cases contain
internal rotations, likewise casting doubt on the applicability of the
RRHO approximation for the zero-point energy. (For both B$_2$H$_4$ and Si$_2$H$_6$, W2 
calculations were feasible, and were found to yield essentially the same result as the
W1 calculation.) Neither neutral nor cationic systems exhibit appreciable
multireference character which could negatively affect the quality of the
W1 and W2 results.
Upon eliminating the four doubtful species, we find a MAD for the G2-2 test
set that is only slightly higher than for the G2-1 set. Regardless of
whether these four species are eliminated, it is clear that W1 
represents a significant improvement over G2 and G3 theory.

\subsection{Heats of Formation}

We shall finally turn to heats of formation for a larger set of 
molecules than was considered in the original W1/W2 paper. 
A comparison with experiment and more approximate methods can be found in 
Table \tabhofdgeetwooneref, while a breakdown
by components of the atomization energies of the G2-1 set of neutral
molecules is given in Table \tabhofgeetwooneref.

First of all, we note that the mean uncertainty for the experimental
values is itself 0.6 kcal/mol. In fact, the MAD values for W1 and W2 theory
stand at 0.6 and 0.5 kcal/mol, respectively, suggesting that these
theoretical values are in the same reliability range as the 
experimental data. The MAD for W2 theory is more than twice as large 
as in the original W1/W2 paper: however, comparisons there were made
against a smaller set of molecules for which the experimental 
uncertainties were all 0.25 kcal/mol or less, mostly 0.1 kcal/mol or
less.

For ten species do discrepancies between W1 theory and 
experiment reach or exceed 1 kcal/mol. Out of these, four experimental 
values carry uncertainties of 1 kcal/mol or more, and can be ignored.
Of the remaining six, the experimental heat of formation of SiH$_{4}$
contains an ambiguity\cite{Mar99SiH4}, P$_{2}$ is a notoriously 
difficult molecule\cite{PerssonTaylorP4} and carries an uncertainty of 
0.5 kcal/mol, and ClO is strongly multireference and carries an uncertainty
of 0.5 kcal/mol. (For this latter molecule, however, `upgrading' the
calculation to W2 theory reduces the discrepancy with experiment to 0.5
kcal/mol, suggesting that slow basis set convergence may be at stake here.)
As for Si$_{2}$H$_{6}$, W1, W2, G2, and CBS-Q theories
exhibit similar discrepancies from experiment (G3 a somewhat smaller 
one), strongly suggesting that the experimental value may be in error.

We note
that W1/W2 and the less expensive methods ``err'' in the same 
direction for P$_{2}$ and ClO as well, suggesting that some revision 
of the experimental data may be in order there as well. Revisions for
BeH and NH$_{2}$ were suggested previously\cite{Martin1997hydrides}:
all methods unanimously suggest the PH$_{2}$ value to be in error.

As pointed out repeatedly\cite{fritz}, the JANAF heat of formation for
SiH$_{4}$ is in fact the older Gunn and Green value\cite{GunnGreen}
increased by a somewhat arbitrary term of 1 kcal/mol for
a Si(amorphous)$\rightarrow$Si(cr) phase change. The W1 and W2 
results, like a previous benchmark study\cite{sih4}, favor the
older Gunn-Green value.

Finally, we selected 26 species with relatively small error bars 
out of the 93 molecules in the G2-2 test set for heats of formation.
The experimental \hof\ along with deviation of W1, G2, and G3 values from
experiment are presented in Table \tabhofdgeetwotworef, while a summary of our
computed TAEs and their different components for the 27 G2-2 neutral test molecules set
calculated at the W1 level is presented in Table \tabhofgeetwotworef.

The average discrepancy between the W1 values and experiment for this
subset is 0.7 kcal/mol, compared to an average experimental uncertainty 
of 0.4 kcal/mol. In order to establish the reason for some of the discrepancies,
we have carried out W2 calculations on selected systems.

In the case of BF$_3$ and CF$_4$, the culprit appears to be slow basis set
convergence in these highly ionic systems. We were unable to complete the
CCSD/aug-cc-pV5Z calculation for CF$_4$: but applying the W1 and W2 extrapolations
to the published CCSD(T)/aug-cc-pV$n$Z ($n$=3,4 and 4,5, respectively) total energies
for CF$_4$ of Dixon et al.\cite{DixonCF4}, we found the estimated W2 TAE to be 1.5 kcal/mol
lower than the estimated W1 TAE. This accounts for essentially all the discrepancy
between experiment and W1 theory for CF$_4$. (A similar phenomenon was previously noted
for BF$_3$\cite{bf3cwb}.) The NO$_2$ molecule exhibits strong nondynamical correlation
effects, and the W2 result is actually further removed from experiment than the
W1 result. ClNO likewise exhibits substantial nondynamical correlation, and here
the W2 result is basically identical to its W1 counterpart.
Improving agreement with experiment for these two molecules will certainly
require accounting for correlation effects neglected at the CCSD(T) level.
While this may be true to a lesser extent of the N$_2$O molecule, an error in 
the experimental value cannot completely be ruled out there.

For F$_2$O, the discrepancies of -0.8 (W1) and -1.0 (W2) kcal/mol with 
experiment, as well as the absence of significant nondynamical correlation,
suggest that the experimental value may need to be reinvestigated.

At least for some of the larger systems (as well as those which have internal 
rotations), neglect of anharmonicity in the zero-point energy may account
for part of the discrepancy with experiment.

\subsection{Proton affinities}

Curtiss et al.\cite{g3paper}, in the original G3 paper, considered proton affinities
of eight molecules as well. We have calculated W1 and W2 proton affinities
for the same systems. However, rather than the somewhat older experimental data
used by these authors, we have taken the data from the very recent compilation
by Hunter and Lias\cite{LiasPA}. Various methods are compared with experiment 
in Table \tabPAtableref, while a breakdown of the different components at the
W1 level is given in Table \tabPAdecompref.

Since protonation/deprotonation is an isogyric reaction, the proton affinity converges
considerably more rapidly with the level of theory than, say, the heat of formation.
(This is expressed, for instance, in the comparatively small contribution of 
valence correlation, and the quite small contributions of inner-shell correlation
and scalar relativistic effects.)
Indeed, we note that only minute differences exist between the W1 and W2
proton affinities: W1 theory can basically be considered converged for this purpose.
Mean absolute deviation from experiment is 0.43 kcal/mol (compared to 1.2 kcal/mol
for the inexpensive G3 theory). While only a few of the values in Hunter and Lias
carry explicit error bars (e.g. water, $\pm$0.7 kcal/mol, H$_2$S, $\pm$1.3 kcal/mol),
it is clear that the uncertainty on the computed W1 values is considerably lower
than that of the experimental values themselves (with the exception of H$_2$,
for which the Hunter and Lias value is a theoretical one). As such, W1 theory should be
a powerful tool for obtaining benchmark-quality proton affinities: for application
to larger systems, the inner-shell correlation and scalar relativistic steps can
fairly safely be skipped for this application.

The somewhat surprising difference of 0.4 kcal/mol for PA(H$_2$) (after all,
both unprotonated and protonated systems should be treatable essentially exactly
at this level) is in part due to 
the harmonic approximation for the zero-point energy. We have calculated 
CCSD(T)/cc-pVQZ quartic force fields for H$_2$ and H$_3^+$, and found 
that the anharmonic zero-point energies thus obtained lead to PA(H$_2$)=101.10 
kcal/mol, or only 0.2 kcal/mol higher than the evaluated Hunter and Lias value.

\section{Conclusions}

We have assessed the performance of two recently developed methods
for benchmark-quality computational thermochemistry, W1 and W2
theory, for a fairly large set of heats of formation, as well
as for the G2-1 and G2-2 sets of ionization potentials and 
electron affinities, and a number of proton affinities.

For molecules which exhibit slow basis set convergence, the
numerical stability of the W1 method is considerably enhanced by
substituting the three-point geometric extrapolation of the
SCF component, $E_{SCF}(L)=E_{SCF,\infty}+A/B^L$, by a 
two-point extrapolation $E_{SCF}(L)=E_{SCF,\infty}+A/L^5$ which
does not involve results with the smallest of the three valence
basis sets.

W1 theory performs excellently for ionization potentials, 
achieving a mean absolute deviation (MAD) of 0.013 eV for the G2-1
set and 0.018 eV for the G2-2 set (minus CN, B$_{2}$H$_{4}$, 
sec-C$_{3}$H$_{7}$, Si$_2$H$_6$, and CH$_{3}$OF). Both mean and
maximum errors are several times
smaller than those of other, less expensive, theoretical 
thermochemistry methods. The vastly more expensive W2 method 
(which uses the same correlation methods but larger basis
sets) yields only a marginal improvement over the W1 method:
it appears that the main impediments to greater accuracy 
for these properties are
the residual imperfections (higher-order $T_3$, $T_4$)
in the CCSD(T) method.

The performance of W1 theory for electron affinities is similar,
with a MAD of 0.016 eV for the G2-1 and 0.019 eV for the G2-2 set.
The latter value is reduced to 0.016 eV if two strongly 
multireference systems (O$_{3}$ and CH$_{2}$NC) are eliminated.
For the G2-1 set, W2 theory (MAD=0.012 eV) does represent a minor 
improvement over W1 theory, reflecting the stronger basis set
sensitivity of electron affinities.

Inner-shell correlation is somewhat more important for ionization
potentials than for electron affinities: scalar relativistic
effects cannot be neglected for either property if results of the
highest accuracy are desired. For IPs or EAs involving systems
in degenerate states, spin-orbit splitting corrections are essential.
With the exception of ClO (where a fairly large active space
is required for good results), sufficiently accurate spin-orbit
splittings can be computed at the CISD/MTsmall level provided the (2s2p)
orbitals on second-row atoms are included in the correlation.

Comparison with experiment for the heats of formation in the G2-1
and (part of) the G2-2 set is complicated somewhat by the uncertainties
in the experimental values: the MADs of both W1 and W2 theory are
lower than the average experimental uncertainty. For parametrizing
more approximate methods, W1 and especially W2 level heats of 
formation may well be more suitable than the experimental data.

Computed proton affinities at the W1 level appear to be 
converged with the level of theory, and agree excellently with
experiment. 

As a final conclusion, we have established that the previously 
proposed W1 and W2 theories are in fact valuable and powerful
tools for accurate ab initio thermochemistry, with mean and 
maximum absolute errors that are several times smaller than those
of more popular (and less expensive) schemes such as G2, G3, and
CBS-Q/CBS-QB3.

\acknowledgments
SP acknowledges a Postdoctoral Fellowship from the Feinberg 
Graduate School (Weizmann Institute). JM is the incumbent of the Helen 
and Milton A. Kimmelman Career Development Chair. This research was 
supported by the Minerva Foundation, Munich, Germany, and by the 
{\it Tashtiyot} Program of the Ministry of Science (Israel).
The authors would like to thank Profs. Joel F. Liebman (NIST and UMBC)
and Chava Lifshitz (Hebrew U. of Jerusalem) for helpful discussions 
on some of the experimental data.

\section*{Supplementary material} 
Calculated total energies and 
geometries of the species discussed in the paper,
as well as aug-cc-pV$n$Z basis sets for Li, Be, Na, and Mg,
are available on the World Wide Web at the Uniform Resource Locator
(URL) \url{http://theochem.weizmann.ac.il/web/papers/w1w2.html}
as well as in Ref.\cite{epaps}.

\begin{table}  \linespread{1.2}
\caption{\label{scfextrap}
Comparison of different extrapolation procedures for 
the SCF component (kcal/mol)}
\begin{tabular}{lcccccc}
Species  &  \multicolumn{2}{c}{W1}   &\multicolumn{2}{c}{W1$'$}   & \multicolumn{2}{c}{W2}\\
         &  \{D,T,Q\} & \{T,Q\} &  \{D,T,Q\} & \{T,Q\} & \{T,Q,5\} & \{Q,5\}   \\
\hline
\multicolumn{5}{c}{Total Atomization Energy}\\
\hline
LiF                 &  90.05  &  93.76  &  90.05  &  93.76 & 93.53 & 93.51  \\
BeH                 &  50.76  &  50.44  &  50.76  &  50.44 & 50.32 & 50.34   \\
SiH$_2$($^3$B$_1$)  &  108.49  &  108.29  &  109.10  &  108.33 & 108.25 & 108.26        \\
SiH$_3$             &  182.86  &  182.66  &  183.72  &  182.73  & 182.55 & 182.58            \\
SiH$_4$             &  260.24  &  259.95  &  261.69  &  260.03  & 259.83 & 259.86            \\
Si$_2$H$_6$         &  424.69  &  424.37  &  426.24  &  424.50 & 424.14 & 424.19 \\
SO$_2$              &  121.51  &  121.74  &  121.87  &  122.00  & 121.94 & 121.98  \\
BF$_3$              &  374.54 & 374.29 & 374.54 & 374.29 & 374.59 & 374.58    \\
C$_6$H$_6$          & 1045.53  &  1045.15  &  1045.53  &  1045.15  & --- & ---   \\
CH$_3$COCH$_3$      &  736.37  &  736.12  &  736.37  &  736.12 & --- & ---    \\
CH$_3$OCH$_3$       &  597.05  &  596.85  &  597.05  &  596.85 & --- & ---  \\
C$_2$H$_4$O(oxirane)&  470.00  &  469.81  &  470.00  &  469.81 & --- & ---      \\
SO$_3$ (a)          &  159.50  &  159.85  &  159.93 &  160.20  & 159.90 & 159.97\\
\hline
\multicolumn{5}{c}{Ionization potential}\\
\hline
Al                  &  126.73  &  126.91  &  126.72  &  126.91 & 126.87 & 126.91 \\
SiH$_4$             &  235.45  &  235.19  &  236.70  &  235.22 & 235.32& 235.32  \\
H$_2$S($^2$A$_1$)   &  270.91  &  270.89  &  271.02  &  270.89 & 270.83 & 270.81 \\
ClF                 &  274.18  &  274.08  &  274.17  &  274.06 & 274.15& 274.24  \\
He                  &  540.86  & 540.62  & 540.86 & 540.62  & 540.71& 540.62\\
CF$_2$              &  236.27  & 236.16  &    236.27  & 236.16  & --- & --- \\
SiH$_3$             & 173.79  & 173.72 & 174.02 & 173.72  & --- & ---   \\
Si$_2$H$_6$         & 198.47  & 198.17 &  199.53  &  198.18 & --- & --- \\
\hline
\multicolumn{5}{c}{Electron affinity}\\
\hline
Si                  &  21.92  &  22.03  &  21.78  &  22.03   & 22.04 & 22.03 \\
SiH$_3$             &  3.96  &  4.04  &  3.74  &  4.03 & 4.31 & 4.31 \\
PH                  &  -2.54  &  -2.53  &  -2.59  &  -2.53 & -2.50 & -2.50  \\
HS                  &  28.06  &  28.14  &  27.92  &  28.13 & 28.12 & 28.12 \\
Cl$_2$              &  43.28  &  43.16  &  43.27  &  43.14 & 43.21 & 43.22 \\
SO$_2$              & 13.93 & 14.14 & 13.88 & 13.79 & --- & --- \\
\end{tabular}

(a) Not part of G2-1 or G2-2 set. Raw data taken from  Ref.\cite{so3tae}.
\end{table}
\newpage

\begin{table} \linespread{1.1}
\caption{\label{eageetwoone}
Components of W1 computed electron affinities (kcal/mol) of the G2-1 ion test set of molecules.}
\bigskip \setlength{\columnsep}{0.0pt}
\squeezetable
\begin{tabular}{l*{9}{r}}
        & SCF   & CCSD  & (T)     & Core  & spin-orbit & Scalar rel. &  Final  & ZPVE &   W1 \\   
Species & limit & limit & limit   & corr. & splitting  &  effects    & Energy  &      &   EA  \\  
\hline
\\

C       & 12.66  &  14.36  &  1.73  &  0.28  &  -0.08  &  -0.07  &  28.88  &    &  28.88             \\      
O       & -12.41  &  42.03  &  3.81  &  0.14  &  -0.05  &  -0.17  &  33.35  &    &  33.35                  \\      
F       & 30.19  &  44.76  &  4.17  &  0.17  &  -0.38  &  -0.26  &  78.64  &    &  78.64                   \\      
Si      & 22.03  &  9.10  &  1.52  &  -0.24  &  -0.42  &  -0.21  &  31.78  &    &  31.78                  \\      
P       & -10.59  &  25.69  &  1.85  &  -0.06  &  0.28  &  -0.24  &  16.93  &    &  16.93                  \\      
S       & 20.84  &  25.34  &  2.15  &  0.00  &  -0.09  &  -0.29  &  47.95  &    &  47.95                   \\      
Cl      &  58.34  &  24.18  &  2.30  &  0.03  &  -0.84  &  -0.34  &  83.66  &    &  83.66                  \\      
CH      &  8.71  &  16.50  &  1.95  &  0.25  &  -0.04  &  -0.06  &  27.32  &  -0.49  &  27.81              \\      
CH$_2$  &  -22.70  &  33.34  &  3.04  &  -0.20  &  0.00  &  0.02  &  13.49  &  -1.29  &  14.78            \\      
CH$_3$  &  -32.71  &  29.24  &  3.22  &  -0.16  &  0.00  &  0.00  &  -0.40  &  -1.06  &  0.66             \\      
NH      & -32.56  &  36.54  &  3.60  &  0.09  &  0.08  &  -0.09  &  7.66  &  -0.36  &  8.03               \\      
NH$_2$  &  -22.14  &  34.94  &  3.99  &  0.07  &  0.00  &  -0.07  &  16.81  &  -0.63  &  17.44            \\      
OH      & -3.05  &  40.75  &  4.40  &  0.11  &  -0.19  &  -0.14  &  41.90  &  -0.16  &  42.06             \\      
SiH     &  16.81  &  10.62  &  1.51  &  -0.17  &  -0.20  &  -0.19  &  28.38  &  -0.30  &  28.68           \\      
SiH$_2$ &  10.74  &  12.21  &  1.56  &  -0.09  &  0.00  &  -0.18  &  24.25  &  -0.66  &  24.90           \\      
SiH$_3$ &  4.04  &  24.64  &  1.99  &  0.27  &  0.00  &  0.11  &  31.04  &  -0.94  &  31.98              \\      
PH      & -2.53  &  23.71  &  2.13  &  -0.05  &  0.29  &  -0.22  &  23.31  &  -0.21  &  23.52             \\      
PH$_2$  &  4.50  &  22.08  &  2.29  &  -0.04  &  0.00  &  -0.21  &  28.61  &  -0.44  &  29.05             \\      
HS      &  28.14  &  23.61  &  2.47  &  -0.01  &  -0.54  &  -0.27  &  53.41  &  -0.09  &  53.50            \\      
O$_2$   & -15.48  &  22.84  &  2.07  &  0.02  &  0.23  &  -0.15  &  9.53  &  -0.64  &  10.16              \\      
NO      & -14.46  &  13.99  &  0.68  &  0.07  &  -0.17  &  -0.17  &  -0.05  &  -0.76  &  0.72             \\      
CN      &  76.63  &  13.10  &  -0.16  &  0.21  &  0.00  &  -0.03  &  89.75  &  -0.01  &  89.76              \\      
PO      &  14.98  &  10.23  &  0.46  &  -0.15  &  -0.33  &  -0.30  &  24.89  &  -0.32  &  25.20             \\      
S$_2$   &  22.66  &  15.40  &  0.57  &  -0.16  &  0.57  &  -0.32  &  38.72  &  -0.21  &  38.93              \\      
Cl$_2$  &  43.16  &  11.39  &  0.71  &  -0.09  &  0.00  &  -0.16  &  55.00  &  -0.48  &  55.48             \\      
\end{tabular} 
\end{table}
\newpage

\begin{table}\linespread{1.1}
\caption{\label{eadgeetwoone}
Deviation of electron affinities (eV) from experiment for the G2-1 test set. }
\squeezetable
\begin{tabular}{lllrrrrr}
         &  \multicolumn{2}{c}{Expt.$^a$} & \multicolumn{5}{c}{Deviation(experiment $-$ theory)}   \\
Species  &  \multispan2{\hrulefill}  & \multispan5{\hrulefill}                           \\      
         &  EA  &  $\pm$(uncert.) & W1  &  W2  &  G2$^b$  &  G3$^b$  &  CBS-QB3$^c$  \\
\hline
  C         &  1.2629$^d$  &  0.0003      &  0.011  &  0.007  &  0.070  &  0.070  &  0.082       \\
  O         &  1.461122$^d$  &  0.000003  &  0.015  &  0.012  &  0.060  &  0.126  &  0.087       \\
  F         &  3.401190$^d$  &  0.000004  &  -0.009 &  -0.002 &  -0.080  &  0.002  &  0.035      \\
  Si        &  1.38946$^e$  &  0.00006    &  0.011  &  0.010  &  0.030  &  0.011  &  0.039       \\
  P         &  0.7465$^d$  &  0.0003      &  0.012  &  0.015  &  0.110  &  0.035  &  0.030          \\
  S         &  2.077104$^d$  &  0.000001  &  -0.002 &  0.008  &  0.070  &  0.013  &  -0.017        \\
  Cl        &  3.61269$^d$  &  0.00006    &  -0.015 &  0.002  &  0.010  &  0.007  &  -0.065        \\
  CH        &  1.238  &  0.0078  &  0.032  &  0.029  &  0.110  &  0.059  &  0.108            \\      
  CH$_2$    &  0.652  &  0.006  &  0.011  &  0.002  &  -0.010  &  0.071  &  0.030           \\      
  CH$_3$    &  0.08  &  0.03  &  0.051  &  0.034  &  0.040  &  0.119  &  0.091              \\      
  NH        &  0.37  &  0.004             &  0.022  &  0.008  &  0.100  &  0.175  &  0.108              \\
  NH$_2$    &  0.776  &  0.037            &  0.020  &  0.007  &  0.000  &  0.078  &  0.056            \\
  OH        &  1.8277  &  0.000044        &  0.004  &  -0.001 &  -0.040  &  0.050  &  0.061      \\
  SiH       &  1.2771  &  0.0087          &  0.034  &  0.031  &  0.090  &  0.007  &  0.082           \\
  SiH$_2$   &  1.123  &  0.022            &  0.043  &  0.039  &  0.140  &  0.048  &  0.108             \\
  SiH$_3$   &  1.406  &  0.014            &  0.019  &  0.011  &  -0.010  &  -0.021  &  0.043           \\
  PH        &  1.028  &  0.01             &  0.008  &  0.010  &  0.070  &  0.048  &  0.026                \\
  PH$_2$    &  1.271  &  0.01             &  0.011  &  0.013  &  0.020  &  0.000  &  0.004               \\
  HS        &  2.317  &  0.002            &  -0.003 &  0.008  &  0.060  &  -0.003  &  -0.013           \\
  O$_2$     &  0.451  &  0.007            &  0.010  &  -0.003 &  -0.030  &  0.052  &  -0.009            \\
  NO        &  0.026  &  0.005            &  -0.005 &  -0.001 &  0.090  &  0.030  &  0.017             \\
  CN        &  3.862  &  0.005            &  -0.031 &  -0.026 &  -0.110  &  -0.067  &  -0.048          \\
  PO        &  1.092  &  0.01             &  -0.001 &  -0.002 &  0.050  &  -0.057  &  0.039             \\
  S$_2$     &  1.663  &  0.04             &  -0.025 &  -0.018 &  0.010  &  -0.006  &  -0.026              \\
  Cl$_2$    &  2.4  &  0.2                &  -0.006 &  0.004  &  0.010  &  -0.067  &  -0.121                \\
\hline
Mean abs. Err  &    &  0.017  &  0.016  &  0.012  &  0.057  &  0.049  &  0.054     \\
Max. abs. Err  &    &  0.200  &  0.051  &  0.039  &  0.140  &  0.175  &  0.121     \\
\end{tabular}
$^a$ Unless otherwise indicated, experimental values are those from \cite{webbook}           \\
$^b$ G2 and G3 values from \cite{g3paper}    \\
$^c$ CBS-QB3 values from \cite{cbs3}          \\
$^d$ {\it CRC Handbook of Chemistry and Physics}, 78th ed. (CRC, Boca Raton, FL, 1997).      \\
$^e$ J. Thogersen, L.D. Steele, M. Scheer, C.A. Brodie, and H.K. Haugen, \jpB{29}{1323}{1996}
\end{table}
\newpage

\begin{table}   \linespread{1.1}
\caption{\label{so}
Calculated and experimental spin-orbit contributions (eV).}
\begin{tabular}{lcccc}
         &  \multicolumn{3}{c}{Calc. spin-orbit splitting}  & Spin-orbit splitting    \\
     &  \multispan3{\hrulefill}  &  from experimental                 \\
      & HF   &  CISD$^a$ &  CISD+subval$^b$    &     experimental fine structure   \\
\hline
OH                  & 0.00848  &  0.00833   &  0.00838  &  0.00863                       \\
CH                  & 0.00165  &  0.00162   &  0.00164  &  0.00173                       \\
SH                  & 0.02177  &  0.02104   &  0.02360  &  0.02337              \\
NO                  & 0.00692  &  0.00729   &  0.00732  &  0.00743    \\
OF                  & 0.01067  &  0.01117   &  0.01117  &   ---          \\     
ClO$^c$                 & 0.01822  &  0.01801   &  0.01917  &  0.01971        \\
ClO                 & 0.01223  &  0.01391   &  0.01378  &          \\
NCCN$^+$            & 0.00305  &  0.00333   &  0.00333  &  ---          \\
CS$_2^+$          & 0.02343  &  0.02437   &  0.02678  &  ---          \\
OCS$^+$             & 0.02198  &  0.02157   &  0.02381  &   ---          \\
CO$_2^+$          & 0.00876  &  0.00936   &  0.00939  &   ---          \\
N$_2^+$($^2$$\Pi$)& 0.00415  &  0.00439   &  0.00439  &  0.00463  \\
HCCH$^+$            & 0.00172  &  0.00183   &  0.00183  &  ---          \\
NH$^+$              & 0.00480  &  0.00475   &  0.00480  &  0.00482     \\
PH$^+$              & 0.01678  &  0.01605   &  0.01852  &  ---          \\
ClF$^+$             & 0.03911  &  0.03810   &  0.04174  &  0.03906    \\
HF$^+$              & 0.01761  &  0.01732   &  0.01740  &  0.01815       \\
HCl$^+$             & 0.03742  &  0.03627   &  0.03999  &  0.04018      \\
Cl$_2^+$          & 0.03965  &  0.04031   &  0.04416  &  0.03998     \\
O$_2^+$           & 0.01181  &  0.01199   &  0.01199  &  0.01223      \\
P$_2^+$           & 0.01328  &  0.01365   &  0.01521  &  0.01612      \\
S$_2^+$           & 0.02649  &  0.02649   &  0.02970  &  0.02914       \\
NH$^{-}$          & 0.00344  &  0.00341   &  0.00344  &  ---\\
O$_{2}^{-}$       & 0.00941  &  0.00976   &  0.00976  &  0.00992\\
SiH               & 0.00787  &  0.00748   &  0.00882  &  0.00885\\
PO                & 0.01293  &  0.01275   &  0.01426  &  0.01389\\
PH$^{-}$          & 0.01117  &  0.01086   &  0.01241  &  ---\\
S$_{2}^{-}$       & 0.02169  &  0.02214   &  0.02455  &  (0.026)\\
\end{tabular}
Experimental fine structure data from Ref.\cite{Hub79}\\
$^a$  CISD with valence correlation only           \\
$^b$  CISD with valence correlation, and 2$s$2$p$ correlation in second-row atoms    \\
$^c$  CASSCF calculated values                \\
\end{table}
\newpage

\begin{table}  \linespread{1.1}
\caption{\label{eadgeetwotwo}
Deviation of electron affinities (eV) from experiment for the G2-2 test set. }
\squeezetable
\begin{tabular}{lllrrr}
         &  \multicolumn{2}{c}{Expt.$^a$} & \multicolumn{3}{c}{Deviation(experiment $-$ theory)}   \\
Species  &  \multispan2{\hrulefill}  & \multispan3{\hrulefill}  \\
         &  EA  &  $\pm$(uncert.) & W1  &  G2$^b$  &  G3$^b$   \\
\hline
Li  &  0.61759  &  0.00022  &        0.001$^c$  &  -0.132  &  -0.124                      \\
B  &  0.27972$^d$  &  0.00003  &       0.021    &  0.090  &  0.076                        \\
Na & 0.547951    &  0.000044  &  -0.002$^c$ & -0.132  &  -0.159                    \\
Al  &  0.43283$^e$  &  0.00005  &      0.020    &  0.083  &  0.043                       \\
C$_2$  &  3.273  &  0.008  &  0.010  &  0.173  &  0.116                         \\
C$_2$O  &  2.289  &  0.018  &          -0.016   &  -0.041  &  -0.001                     \\
CF$_2$  &  0.179  &  0.005  &          0.019    &  0.089  &  0.001                       \\
NCO  &  3.609  &  0.005  &             -0.021   &  -0.011  &  0.032                      \\
NO$_2$  &  2.273  &  0.005  &  -0.014  &  -0.067  &  -0.008                     \\
O$_3$  &  2.103  &  0.004  &           -0.066   &  0.033  &  0.000                        \\
OF  &  2.272  &  0.006  &              -0.041   &  -0.028  &  0.021                       \\
SO$_2$  &  1.107  &  0.008  &          -0.029   &  -0.053  &  -0.077                     \\
S$_2$O  &  1.877  &  0.008  &          -0.026   &  -0.043  &  -0.113                     \\
CCH  &  2.969  &  0.006  &             0.007    &  -0.121  &  -0.027                     \\
C$_2$H$_3$  &  0.667  &  0.024  &      -0.007   &  -0.083  &  0.012                    \\
CH$_2$CC  &  1.794  &  0.008  &  -0.024  &  0.054  &  -0.027                    \\
CH$_2$CCH  &  0.893  &  0.005  &       -0.016   &  -0.097  &  -0.013           \\
CH$_2$CHCH$_2$  &  0.481  &  0.008  &  -0.005   &  -0.039  &  0.039                  \\
HCO  &  0.313  &  0.005  &             0.000    &  -0.027  &  0.005                         \\
HCF  &  0.542  &  0.005  &             0.004    &  0.082  &  0.013                           \\
CH$_3$O  &  1.570  &  0.005  &         0.014    &  -0.050  &  0.017                   \\
CH$_3$S  &  1.861  &  0.004  &  -0.010  &  -0.009  &  0.001                        \\
CH$_2$S  &  0.465  &  0.023  &         -0.041   &  0.075  &  0.001                         \\
CH$_2$CN  &  1.544$^f$  &  0.006  &    -0.002   &  -0.036  &  0.026                        \\
CH$_2$NC  &  1.058$^f$  &  0.026  &    -0.075   &  -0.122  &  -0.048                 \\
CHCO  &  2.35  &  0.022  &  0.008  &  -0.010  &  0.039                          \\
CH$_2$CHO  &  1.824157  &  0.000044  & -0.008   &  -0.046  &  -0.010               \\
CH$_3$CO  &  0.423  &  0.037  &        0.027    &  -0.017  &  0.020                        \\
CH$_3$CH$_2$O  &  1.713  &  0.005  &  -0.047  &  -0.097  &  -0.039              \\
LiH  &  0.342  &  0.012  &             0.013    &  0.022  &  -0.053                          \\
HNO  &  0.338  &  0.015  &             0.003    &  0.088  &  0.043                          \\
HO$_2$  &  1.078  &  0.017  &          -0.002   &  -0.032  &  0.029                        \\
\hline
Mean abs. Err    &    &  0.010  &  0.019  &  0.065  &  0.039                                      \\ 
Max. abs. Err    &    &  0.037  &  0.075  &  0.173  &  0.159 
\end{tabular}
$^a$ Unless otherwise indicated, experimental values are those from \cite{webbook}           \\
$^b$ G2 and G3 values from \cite{g3paper}    \\
$^c$ augmented basis sets were used        \\
$^d$ M. Scheer, R.C. Bilodeau, and H.K. Haugen, \PRL{80}{2562}{1998}    \\
$^e$ M. Scheer, R.C. Bilodeau, J. Thogersen, and H.K. Haugen, \PRA{57}{1493}{1998} \\
$^f$ Ref. \cite{Ber94}
\end{table}
\newpage

\begin{table}  \linespread{1.1}
\caption{\label{ipgeetwoone}
Components of W1 computed ionization potentials (kcal/mol) of the G2-1 ion test set of molecules.}
\squeezetable
\begin{tabular}{l*{9}{r}}
        & SCF   & CCSD  & (T)     & Core  & spin-orbit & Scalar rel. &  Final  & ZPVE &   W1 \\
Species & limit & limit & limit   & corr. & splitting  &  effects    & Energy  &      &   IP  \\
\hline
\\
Li      &  123.18  &  0.00  &  0.00  &  1.02  & 0.00   &  0.01  &  124.21  &    &  124.21                       \\
Be      &  185.51  &  29.03  &  0.00  &  0.40  &  0.00  &  0.02  &  214.97  &    &  214.97                 \\
B  &  182.94  &  6.08  &  1.05  &  0.88  &  0.03  &  -0.05  &  190.93  &    &  190.93           \\
C  &  248.73  &  9.29  &  0.74  &  0.76  &  -0.04  &  -0.09  &  259.40  &    &  259.40          \\
N  &  321.79  &  12.29  &  0.67  &  0.67  &  -0.23  &  -0.14  &  335.06  &    &  335.06         \\
O  &  275.90  &  36.06  &  1.30  &  0.33  &  0.22  &  -0.20  &  313.62  &    &  313.62          \\
F  &  362.02  &  37.92  &  1.61  &  0.34  &  -0.09  &  -0.29  &  401.51  &    &  401.51         \\
Na  &  114.18  &  0.00  &  0.00  &  3.95  &  0.00  &  0.15  &  118.28  &    &  118.28           \\
Mg  &  152.37  &  21.36  &  0.00  &  2.07  &  0.00  &  0.22  &  176.02  &    &  176.02          \\
Al  &  126.91  &  9.93  &  0.97  &  -0.33  &  0.21  &  -0.22  &  137.48  &    &  137.48         \\
Si  &  176.45  &  10.32  &  1.14  &  -0.08  &  -0.09  &  -0.25  &  187.49  &    &  187.49       \\
P  &  231.39  &  9.90  &  1.24  &  0.08  &  -0.85  &  -0.28  &  241.48  &    &  241.48          \\
S  &  209.63  &  27.22  &  1.22  &  0.23  &  0.56  &  -0.32  &  238.54  &    &  238.54          \\
Cl  &  271.73  &  26.08  &  1.36  &  0.25  &  -0.13  &  -0.36  &  298.94  &    &  298.94        \\
CH$_4$(D$_{2d}$ cation)  & 278.37  &  18.94  &  0.68  &  0.14  &  0.00  &  -0.03  &  298.10  &  5.43  &  292.67    \\
CH$_4$(C$_{2v}$ cation)$^a$  &  271.66  &  22.52  &  1.05  &  0.27  &  0.00  &  -0.03  &  295.47  &  3.93  &  291.54 \\
NH$_3$  &  202.33  &  31.12  &  2.35  &  0.09  &  0.00  &  -0.02  &  235.87  &  1.04  &  234.83            \\
OH     &  264.15  &  34.73  &  1.88  &  0.28  &  0.19  &  -0.16  &  301.07  &  0.93  &  300.14      \\
H$_2$O  &  255.86  &  34.37  &  2.39  &  0.24  &  0.00  &  -0.13  &  292.73  &  1.68  &  291.05     \\
FH     &  331.78  &  38.03  &  2.28  &  0.27  &  0.40  &  -0.24  &  371.72  &  1.54  &  370.18      \\
SiH$_4$  &  235.19  &  19.71  &  0.61  &  0.23  &  0.00  &  -0.25  &  255.49  &  1.71  &  253.78    \\
PH     &  221.61  &  11.75  &  1.25  &  0.17  &  -0.43  &  -0.25  &  234.11  &  -0.01  &  234.12    \\
PH$_2$  &  211.53  &  13.52  &  1.33  &  0.26  &  0.00  &  -0.23  &  226.42  &  -0.03  &  226.45    \\
PH$_3$  &  198.75  &  26.57  &  1.89  &  0.49  &  0.00  &  0.19  &  227.89  &  0.21  &  227.67      \\
HS     &  213.11  &  25.26  &  1.54  &  0.21  &  0.54  &  -0.29  &  240.37  &  0.20  &  240.17       \\
H$_2$S($^2B_1$)  &  216.34  &  23.62  &  1.80  &  0.19  &  0.00  &  -0.27  &  241.67  &  0.39  &  241.28     \\
H$_2$S($^2A_1$)  &  270.89  &  22.49  &  1.40  &  0.30  &  0.00  &  0.24  &  295.31  &  0.39  &  294.91      \\
HCl  &  269.43  &  24.27  &  1.71  &  0.21  &  -0.92  &  -0.33  &  294.37  &  0.45  &  293.92      \\
C$_2$H$_2$  &  227.74  &  31.98  &  3.30  &  0.64  &  -0.04  &  -0.04  &  263.59  &  0.61  &  262.98      \\
C$_2$H$_4$  &  207.05  &  33.10  &  3.40  &  0.49  &  0.00  &  -0.04  &  244.01  &  1.44  &  242.56      \\
CO  &  306.44  &  16.79  &  -0.20  &  0.22  &  0.00  &  0.02  &  323.27  &  -0.11  &  323.38           \\
N$_2$($^2\Sigma^+$cation)  &  368.35  &  -4.69  &  -3.62  &  0.56  &  0.00  &  -0.07  &  360.52  &  0.17  &  360.36 \\
N$_2$($^2\Pi$cation)  &  354.09  &  29.27  &  1.98  &  0.96  &  -0.10  &  -0.16  &  386.03  &  0.17  &  385.87 \\
O$_2$ &  271.54  &  7.58  &  -0.81  &  0.32  &  -0.28  &  -0.21  &  278.15  &  -0.59  &  278.75         \\
P$_2$ &  214.65  &  24.92  &  3.35  &  0.21  &  -0.35  &  -0.36  &  242.42  &  0.06  &  242.36          \\
S$_2$ &  208.84  &  8.59  &  -0.61  &  0.04  &  -0.68  &  -0.39  &  215.77  &  -0.17  &  215.94         \\
Cl$_2$ &  255.59  &  11.60  &  -0.79  &  0.09  &  -1.02  &  -0.40  &  265.07  &  -0.13  &  265.20       \\
ClF    &  274.08  &  18.66  &  0.13  &  0.19  &  -0.96  &  -0.32  &  291.78  &  -0.20  &  291.98        \\
CS    &  243.16  &  17.48  &  0.91  &  0.01  &  0.00  &  -0.06  &  261.51  &  -0.15  &  261.66          \\
\end{tabular}
$^a$ geometry optimized at CCSD(T)/cc-pVTZ level. B3LYP/cc-pVTZ optimization erroneously yields
$D_{2d}$ structure (see text and Table \tabmethaneplusfreqref).
\end{table}
\newpage

\begin{table}  \linespread{1.1}
\caption{\label{ipdgeetwoone}
Deviation of ionization potentials (eV) from experiment for the G2-1 test set. }
\squeezetable
\begin{tabular}{lllrrrrr}
         &  \multicolumn{2}{c}{Expt.$^a$} & \multicolumn{5}{c}{Deviation(experiment $-$ theory)}   \\
Species  &  \multispan2{\hrulefill}  & \multispan5{\hrulefill}  \\
         &  IP  &  $\pm$(uncert.) & W1  &  W2  &  G2$^b$  &  G3$^b$  &  CBS-QB3$^c$  \\
\hline
Li        & 5.39172 &  0.00001  &               0.005   &  0.005  &  0.082  &  -0.007  &  0.084         \\
Be        &  9.32263  &  0.00001  &           0.001  &  0.005  &  -0.090  &  -0.135  &  -0.048              \\
B         &  8.29802  &  0.00002  &             0.019   &  0.007  &  0.100  &  0.063  &  0.074          \\
C         &  11.2603  &  0.0001  &              0.012   &  0.010  &  0.080  &  0.051  &  0.070          \\
N         &  14.534  &  0.001  &              0.004  &  0.000  &  0.060  &  0.029  &  0.056              \\
O         &  13.618  &  0.001  &                0.018   &  0.005  &  0.080  &  0.071  &  0.022          \\
F         &  17.423  &  0.001  &                0.012   &  0.002  &  0.030  &  0.034  &  -0.052         \\
Na        &  5.139  &  0.001  &               0.010  &  0.010  &  0.189  &  0.027  &  0.009                \\
Mg        &  7.646  &  0.001  &               0.013  &  0.013  &  -0.004  &  -0.137  &  0.053              \\
Al        &  5.986  &  0.001  &                 0.024   &  0.023  &  0.050  &  0.028  &  0.061           \\
Si        &  8.15166 &   0.00003  &            0.021  &  0.018  &  0.050  &  0.025  &  0.060               \\
P         &  10.48669  &  0.00001  &          0.015  &  0.011  &  0.037  &  0.023  &  0.079              \\
S         &  10.360 &  0.001  &                 0.016   &  0.014  &  0.160  &  0.092  &  0.091             \\
Cl        &  12.968  &  0.001  &                0.005   &  0.007  &  0.120  &  0.076  &  0.047            \\
CH$_4$(D$_{2d}$ cation)     &  12.61  &  0.01  &  -0.081  &  -0.082  &  -0.060  &  -0.043  &  -0.100              \\
CH$_4$(C$_{2v}$ cation)$^d$ &  12.61  &  0.01  &    -0.032  &  -0.033  &  -0.060  &  -0.043  &  -0.100       \\
NH$_3$    &  10.18  &  0.09  &  -0.003  &  -0.004  &  0.101  &  0.042  &  0.115             \\
OH        &  13.017  &  0.0002  &                  0.002   &  0.001  &  0.030  &  0.082  &  -0.013          \\
H$_2$O    &  12.6223  &  0.0003  &  0.001  &  0.006  &  -0.010  &  0.030  &  -0.039             \\
FH        &  16.044  &  0.003  &                   -0.008  &  -0.016  &  -0.050  &  0.000  &  -0.091       \\
SiH$_4$   &  11.0  &  0.02  &                      -0.005  &  0.006  &  -0.010  &  -0.023  &  -0.009       \\
PH        &  10.149  &  0.008  &                   -0.003  &  -0.006  &  0.060  &  -0.037  &  0.043        \\
PH$_2$    &  9.824  &  0.002  &                    0.004   &  0.003  &  0.100  &  0.007  &  0.061           \\
PH$_3$    &  9.87  &  0.002  &                     -0.003  &  -0.006  &  0.000  &  -0.012  &  0.026        \\
HS        &  10.4219  &  0.0004  &                 0.007   &  0.007  &  0.060  &  0.097  &  0.026          \\
H$_2$S($^2B_1$)  &  10.453  &  0.001  &          -0.010  &  -0.008  &  0.040  &  0.011  &  0.022        \\
H$_2$S($^2A_1$)  &  12.76  &  0.036  &           -0.029  &  -0.029  &  0.030  &  -0.002  &  -0.004      \\
HCl       &  12.747  &  0.002  &                   0.001   &  0.008  &  0.030  &  0.029  &  -0.026        \\
C$_2$H$_2$   &  11.403  &  0.0003  &               -0.001  &  -0.004  &  -0.020  &  -0.006  &  -0.039     \\
C$_2$H$_4$   &  10.5138  &  0.0006  &              -0.005  &  -0.001  &  -0.070  &  -0.045  &  -0.035   \\
CO       &  14.0142  &  0.0003  &                  -0.009  &  -0.014  &  0.000  &  -0.001  &  -0.056       \\
N$_2$ ($^2\Sigma^+$cation) & 15.581  &  0.008 &    -0.046  &  -0.046  &  0.020  &  0.018  &  -0.030        \\
N$_2$($^2\Pi$cation)    & 16.699  &  0.001  &    -0.034  &  -0.049  &  0.030  &  0.030  &  -0.052        \\
O$_2$  &  12.0697  &  0.0002  &                    -0.018  &  -0.024  &  -0.100  &  -0.176  &  -0.095      \\
P$_2$  &  10.567  &  0.002  &                      0.057   &  0.047  &  -0.010  &  0.017  &  -0.013         \\
S$_2$  &  9.356  &  0.002  &  -0.008  &  -0.011  &  0.080  &  -0.023  &  -0.048                     \\
Cl$_2$  &  11.481  &  0.003  &                     -0.019  &  -0.008  &  -0.010  &  -0.045  &  -0.039     \\
ClF     &  12.66  &  0.01  &                       -0.002  &  0.005  &  -0.070  & -0.002  &  -0.110        \\
CS     &  11.33  &  0.01  &                        -0.017  &  -0.017  &  -0.090  &  -0.061  &  0.048        \\
\hline
Mean abs. error.  &    & 0.006 &  0.013  &  0.013  &  0.058  &  0.043  &  0.051                          \\
Max abs. error.  &   &  0.090 &  0.057  & 0.049  & 0.189 & 0.176  & 0.115            \\
\end{tabular}
$^a$ Experimental values from  \cite{webbook}            \\
$^b$ G2 and G3 values from \cite{g3paper}    \\
$^c$ CBS-QB3 values from \cite{cbs3}     \\
$^d$ CCSD(T)/cc-pVTZ geometry. B3LYP/cc-pVTZ optimization erroneously yields $D_{2d}$ structure
(see text and Table \tabmethaneplusfreqref).
\end{table}
\newpage
\begin{table} \linespread{1.5}
\caption{\label{ch4+freq}
Equilibrium structure (\AA, degrees) and harmonic frequencies (cm$^{-1}$) 
of the $C_{2v}$ structure of CH$_4^+$ obtained with the cc-pVTZ basis set
and a variety of electronic structure methods}
\squeezetable
\begin{tabular}{lrrrrrrrrr}
  &  CCSD(T)  &  B3LYP  &  BLYP  &  BHLYP  &  BHPW91  &  B3PW91  &  B3P86  &  mPW1PW91  &  mPW1K   \\
\hline
$r_{CH1}$  &  1.0826  &  1.0827  &  1.0888  &  1.0752  &  1.0764  &  1.0838  &  1.0829  &  1.0825  &  1.0781        \\
$r_{CH2}$  &  1.1864  &  1.1867  &  1.1955  &  1.1766  &  1.1769  &  1.1869  &  1.1858  &  1.1851  &  1.1790        \\
$\theta_{H1CH1'}$  &  125.62  &  124.20  &  124.12  &  124.23  &  124.88  &  124.75  &  124.72  &  124.80  &  124.84    \\
$\theta_{H2CH2'}$  &  55.01  &  57.56  &  58.22  &  57.00  &  55.76  &  56.52  &  56.58  &  56.34  &  55.97             \\
\hline
$B_2$  &  3308  &  3281  &  3208  &  3370  &  3384  &  3292  &  3298  &  3308  &  3362     \\
$A_1$  &  3165  &  3147  &  3078  &  3232  &  3242  &  3154  &  3161  &  3170  &  3221     \\
$A_1$  &  2572  &  2534  &  2460  &  2620  &  2641  &  2552  &  2557  &  2569  &  2621     \\ 
$B_2$  &  2292  &  2188  &  2097  &  2281  &  2353  &  2246  &  2248  &  2267  &  2328      \\
$A_1$  &  1585  &  1545  &  1507  &  1593  &  1615  &  1562  &  1564  &  1572  &  1602      \\
$B_1$  &  1311  &  1296  &  1269  &  1337  &  1331  &  1292  &  1292  &  1297  &  1321      \\
$A_1$  &  1201  &  1159  &  1126  &  1184  &  1226  &  1187  &  1191  &  1196  &  1218      \\
$B_2$  &  894  &  901  &  875  &  933  &  912  &  885  &  889  &  888  &  906              \\
$A_2$  &  469  &  227$i$  &  417$i$  &  358  &  339  &  233$i$  &  256$i$  &  156$i$  &  269         \\
\end{tabular}
\end{table}
\newpage

\begin{table}  \linespread{1.1}
\caption{\label{ipdgeetwotwo}
Deviation of ionization potentials (eV) from experiment for the G2-2 test set. }
\squeezetable
\begin{tabular}{lllrrr}
         &  \multicolumn{2}{c}{Expt.$^a$} & \multicolumn{3}{c}{Deviation(experiment $-$ theory)}   \\
Species  &  \multispan2{\hrulefill}  & \multispan3{\hrulefill}  \\
         &  IP  &  $\pm$(uncert.) & W1  &  G2$^b$  &  G3$^b$   \\
\hline
H  &  13.599  &  0.001  &  -0.007  &  0.004  &  0.035                                        \\ 
He  &  24.588  &  0.001  &                     0.000   &  -0.048  &  -0.018                                     \\
Ne  &  21.565  &  0.001  &                     -0.012  &  0.048  &  0.013                                     \\
Ar  &  15.760  &  0.001  &                     -0.005  &  -0.067  &  -0.071                                     \\
BF$_3$  &  15.71  &  0.1  &                    -0.028  &  -0.099  &  -0.103                                     \\
CO$_2$  &  13.778  &  0.002  &                 -0.039  &  -0.071  &  -0.079                                 \\ 
CF$_2$  &  11.42  &  0.01  &                   -0.010  &  0.024  &  0.019                                      \\
OCS  &  11.185  &  0.002  &                    0.006   &  -0.023  &  -0.006                                 \\ 
CS$_2$  &  10.08  &  0.002  &                  -0.003  &  0.020  &  0.015                                    \\ 
CH$_2$  &  10.396  &  0.003  &                 0.022   &  -0.084  &  -0.002                                   \\
CH$_3$  &  9.837  &  0.005  &  -0.003  &  -0.054  &  0.033                                    \\
C$_2$H$_5$($^2$A$'$)  &  8.117  &  0.008  &  -0.009  &  -0.047  &  0.048                                   \\
C$_3$H$_4$(cyclopropene)  &  9.668  &  0.005  &  0.032  &  0.093  &  0.067                              \\
CH$_2$CCH$_2$  &  9.6878  &  0.002  &          -0.014  &  0.047  &  0.008                                \\
sec-C$_3$H$_7$  &  7.37  &  0.02  &  -0.079  &  0.028  &  0.093                                 \\ 
C$_6$H$_6$  &  9.24384  &  0.00006  &          -0.001  &  0.084  &  0.066                                \\ 
CN($^1\Sigma^+$ cation)$^c$  &  14.03  &  0.02  &       0.165  &   0.297  &   0.171      \\
CN($^3\Pi$ cation)$^c$  &  14.03  &  0.02  &      -0.039  &  ---     & ---          \\
CHO  &  8.14  &  0.04  &                       -0.010  &  -0.040  &  0.021                                     \\ 
CH$_2$OH($^2$A)  &  7.56  &  0.01  &           0.016   &  -0.110  &  -0.028                                    \\
CH$_3$O($^2$A$'$)  &  10.72  &  0.01  &  -0.006  &  0.060  &  0.021                                     \\
CH$_3$OH  &  10.85  &  0.03  &  -0.035  &  0.116  &  0.077                                  \\
CH$_3$F  &  12.54  &  0.01  &                  -0.021  &  0.161  &  0.144                                     \\
CH$_2$S &  9.376  &  0.003  &                  -0.013  &  -0.001  &  -0.018                                  \\
CH$_2$SH  &  7.536  &  0.003  &                -0.026  &  -0.116  &  -0.034                                 \\
CH$_3$SH  &  9.443  &  0.002  &                -0.007  &  0.019  &  0.015                                   \\
CH$_3$Cl  &  11.265  &  0.003  &               -0.022  &  0.036  &  0.027                                  \\
CH$_3$CHO  &  10.227  &  0.005  &  -0.031  &  0.081  &  0.046                                 \\
CH$_3$OF  &  11.34  &  0.008  &                -0.078  &  0.065  &  0.060                                   \\
C$_2$H$_4$S(thiirane)  &  9.051 & 0.006 &      -0.013  &  0.016  &  0.016                                   \\
NCCN  &  13.374  &  0.008  &                   -0.046  &  0.017  &  0.026                                  \\ 
C$_4$H$_4$O(furan)  &  8.91  &  0.01  &        0.032   &  -0.003  &  -0.020                              \\
B$_2$H$_4$  &  9.70  &  0.02  &  0.129$^d$  &       0.128   &  -0.091                                     \\  
NH  &  13.49  &  0.01  &  0.019  &  -0.077  &  0.005                                      \\ 
NH$_2$  &  11.14  &  0.01  &                   -0.033  &  0.035  &  -0.021                                     \\
N$_2$H$_2$  &  9.589  &  0.007  &              -0.010  &  0.129  &  0.086                                    \\
N$_2$H$_3$  &  7.61  &  0.01  &  -0.005  &  -0.069  &  0.005                                     \\
HOF  &  12.71  &  0.01  &                      -0.022  &  -0.004  &  0.009                                     \\
SiH$_2$($^1$A$_1$)  &  9.15  &  0.02  &        0.015   &  0.022  &  0.043                                       \\
SiH$_3$  &  8.135  &  0.005  &                 0.022   &  -0.082  &  0.017                                   \\ 
Si$_2$H$_2$  &  8.2  &  0.02  &                0.008   &  0.078  &  0.057                                     \\ 
Si$_2$H$_4$  &  8.09  &  0.03  &               0.030   &  0.032  &  0.054                                     \\
Si$_2$H$_6$  &  9.74  &  0.02  &               0.091$^e$    &  -0.035  &  -0.044                                  \\ 
\hline
Mean abs. Err  &    &   0.011 &  0.025  &  0.062  &  0.044                                  \\
 w/o  B$_2$H$_4$ &  &   0.011 &  0.022  &  0.062  &  0.043                            \\
w/o  B$_2$H$_4$, sec-C$_3$H$_7$, CH$_3$OF, Si$_2$H$_6$  &    & 0.010   &  0.017  &  0.063  &  0.041             \\
Max abs. Err    &  & 0.100    &  0.129  & 0.161  &  0.144         \\
w/o  B$_2$H$_4$, sec-C$_3$H$_7$, CH$_3$OF, Si$_2$H$_6$  &    & 0.100   &  0.046  &  0.161  &  0.144             \\
\end{tabular}
{\footnotesize
$^a$ Experimental values from   \cite{webbook}           \\
$^b$ G2 and G3 values from \cite{g3paper}    \\
$^c$ Ionization potential of 13.60 eV given in Ref.\cite{webbook} is propagated
transcription error (see text). Measured value of 14.03$\pm$0.02 eV\cite{Ber69}
corresponds to unspecified state of cation: may be $^3\Pi$ for symmetry reasons.
Consequently, CN has been excluded from the error statistics for all methods.\\
$^d$ W2 theory: IP=9.57 eV, experiment$-$theory=0.127 eV\\
$^e$ W2 theory: IP=9.66 eV, experiment$-$theory=0.084 eV
}
\end{table} 
\begin{table}  \linespread{1.0}
\caption{\label{hofdgeetwoone}
Deviation of heats of formation (kcal/mol) from experiment for the G2-1 test set. }
\squeezetable
\begin{tabular}{lllcrrrrr}
         &  \multicolumn{3}{c}{Expt.$^a$} & \multicolumn{5}{c}{Deviation(experiment $-$ theory)}   \\
Species  &  \multispan3{\hrulefill}  & \multispan5{\hrulefill}  \\
         &  \hof\  &  $\pm$(uncert.) & Ref. & W1  &  W2  &  G2$^b$  &  G3$^b$  &  CBS-Q$^c$  \\
\hline
LiH  &  33.61  &  0.01  & J                     &  0.4 &    0.3                          &  0.9  &  0.6  &  -0.1        \\
BeH  &  81.70  &  1.00  &  H                    &  0.9 &   0.8                          &  -1.5  &  -0.5  &  -0.8      \\
CH  &  142.77  &  0.31  &  G                    &  0.4 &   0.4 			 &  0.9  &  1.7  &  0.2        \\
CH$_2$($^3$B$_1$)  &  93.31  &  0.96 & G        &  0.0 &   -0.1            &  -1.4  &  0.9  &  -1.4     \\
CH$_2$($^1$A$_1$)  & 102.31  &  1.00  & E       & -0.2 &    -0.3            &  0.9  &  0.5  &  -0.6      \\
CH$_3$  &  34.97  &  0.12  &  G                 &  0.4 &   0.2 			 &  -0.1  &  1.0  &  -0.3      \\
CH$_4$  & -17.83  &  0.07  &  G                 &  0.4 &   0.2 			 &  0.8  &  0.4  &  -0.1       \\
NH  & 85.67  &  2.39  &  G                      &  -0.1&    -0.1 		 &  -0.6  &  1.4  &  -0.8      \\
NH$_2$  &  45.50  &  1.51  & J                  &  1.3 &   1.1			  &  0.5  &  1.0  &  0.0       \\
NH$_3$  &  -10.98  &  0.08  & C                 &  0.3 &   0.0  			&  -0.2  &  -0.8  &  -1.0      \\
OH  &  9.40  &  0.05  &  G                      &  0.6 &   0.6 				  &  0.3  &  1.0  &  0.4       \\
OH$_2$  &  -57.80  &  0.01  & C                 &  0.5 &   0.2                       &  0.3  &  -0.3  &  0.1       \\
FH  &   -65.32  &  0.17  &  C                   &  0.4 &   0.0                           &  0.8  &  0.1  &  0.7      \\
SiH$_2$($^1$A$_1$)  &  65.33  &  1.20 & G       &  1.3 &   1.3         &  3.0  &  2.2  &  2.4      \\
SiH$_2$($^3$B$_1$)  &  86.20  &  1.00  & B      &  0.8 &   0.5         &  0.5  &  1.3  &  1.5      \\
SiH$_3$  &  47.90  &  2.39  & DO                &  1.1 &   0.9                  &  1.2  &  1.0  &  2.2       \\
SiH$_4$  & 8.29  &  0.36 & G                    &  1.4 &   1.1                     &  2.3  &  1.0  &  2.8       \\
PH$_2$  &  30.10  &  1.00  &  J                 & -1.8 &    -1.9               &  -2.8  &  -2.5  &  -1.6     \\
PH$_3$  &  1.29  &  0.41 & J                    & -0.3 &   -0.3                   &  -0.7  &  -1.8  &  0.0      \\
H$_2$S  &  -4.92  &  0.12 & C                   &  0.4 &   0.4                    &  -0.1  &  -0.4  &  0.7    \\
HCl  &  -22.06  &  0.02  & C                    &  0.1 &   0.1                      &  0.4  &  -0.2  &  0.9     \\
Li$_2$  &  51.60  &  0.72  & J                  &  0.0 &  -0.1                      &  2.0  &  2.2  &  0.0      \\
LiF  &  -81.45  &  2.01  &  J                   &  0.9 &   0.5                          &  -0.1  &  -0.7  &  -1.0    \\
C$_2$H$_2$  &  54.35  &  0.19  & G              & -0.1 &    -0.4                  &  -1.5  &  -0.6  &  -1.9  \\
C$_2$H$_4$  &  12.52  &  0.12  & G              &  0.6 &   0.2                    &  -0.2  &  0.2  &  -1.0     \\
C$_2$H$_6$  &  -20.08  &  0.10  &  G            &  1.1 &   0.7                    &  0.5  &  0.3  &  -0.6     \\
CN  &  105.23  &  1.20  &  G                    & -0.9 &    -0.9                         &  -2.1  &  -1.5  &  -1.9    \\
HCN  &  31.55  &  0.96  & G                     &  0.4 &   0.3                           &  0.3  &  0.2  &  -0.7     \\
CO  &  -26.42  &  0.04  & G                     &  0.1 &   -0.1                            &  1.8  &  0.3  &  0.6     \\
HCO  &  10.04  &  1.20  & G                     &  0.3 &   0.0                              &  0.7  &  0.3  &  0.5    \\
CH$_2$O  &  -25.98  &  0.12  & G                &  0.5 &   0.4                         &  2.0  &  0.6  &  1.0    \\
CH$_3$OH  &  -48.04  &  0.14  & G               &  1.0 &  ---                         &  1.4  &  0.1  &  0.3    \\
N$_2$  &  0.00  &  0.00  & S                    & -0.5 &    -0.5                         &  -1.3  &  -2.1  &  -2.0  \\
H$_2$NNH$_2$  &  22.75  &   0.12  & G           &  0.2 &  ---                   &  -1.0  &  -2.2  &  -2.2  \\
NO  &  21.58  &  0.04  &  J                     & -0.7 &    -0.7	&  0.6  &  -0.2  &  0.5    \\
O$_2$  &  0.00  &  0.00  &  S                   & -0.4 &    -0.7			  &  -2.4  &  -1.1  &  0.2    \\
HOOH  &  -32.48  &  0.05  & G                   &  0.1 &   -0.4 			 &  -0.2  &  -1.2  &  0.2    \\
F$_2$  &  0.00  &  0.00  &  S                   & -0.7 &    -0.8                         &  -0.3  &  -0.7  &  0.6   \\
CO$_2$  &  -94.05  &  0.03  & C                 &  0.4 &   -0.1                      &  2.7  &  1.2  &  2.1     \\
Na$_2$  &  33.96  &  0.29  & J                  &  0.0 &   -0.1                     &  2.4  &  4.0  &  0.4      \\
Si$_2$  &  140.99  &  3.11  &  J                &  0.1 &   0.4                    &  0.7  &  3.0  &  1.0     \\
P$_2$  &  34.42  &  0.48  & C                   & -1.1 &    -0.8			  &  -1.2  &  -1.1  &  -0.3  \\
S$_2$  &  30.74  &  0.07  & C                   &  0.3 &   0.9 			 &  -3.2  &  -0.9  &  1.5      \\
Cl$_2$  &  0.00  &  0.00  & S                   & -0.6 &    0.2 			 &  -1.4  &  -1.1  &  1.7   \\
NaCl  &   -43.36  &  0.50  & J                  & -0.4 &    -0.6 		 &  1.2  &  1.4  &  0.8       \\
SiO  &   -24.00  &  2.01  & J                   & -0.9 &    -0.6  	         &  -1.1  &  -0.1  &  1.8     \\
CS  &  66.86  &  0.23  & H                      & -0.9 &    -0.5                      &  1.0  &  1.1  &  1.3       \\
SO  &  1.20  &  0.31  & J                       & -0.4 &    0.1                         &  -2.6  &  -0.5  &  0.7    \\
ClO  &  24.19  &  0.50  & J                     & -1.3 &    -0.5                      &  -2.2  &  -1.7  &  -0.3   \\
FCl  &  -13.20  &  0.10  & J                    &  0.0 &   0.0                      &  0.7  &  -0.7  &  0.5        \\
Si$_2$H$_6$  &  19.05  &  0.31 & L              &  2.2 &   1.9                &  2.8  &  1.3  &  3.3   \\
CH$_3$Cl  & -19.57  &  0.14  & G                &  0.7 &   0.6                    &  0.9  &  -0.1  &  1.1      \\
CH$_3$SH  &   -5.46  &  0.14  & GO              &  0.8 &  ---                   &  -0.2  &  -0.4  &  0.5     \\
HOCl  & -17.81  &  0.50  & J                    &  0.5 &   0.5                        &  0.5  &  -0.4  &  1.4          \\
SO$_2$  & -70.94  &  0.05 & C                   & -0.7 &    0.2                   &  -4.9  &  -3.8  &  -0.3   \\
\hline
Mean abs. error  &    & 0.54 &                   & 0.6  &  0.5                       &  1.2  &  1.1  &  1.0     \\
Max abs. error  &    & 3.11  &                   & 2.2  &  1.9                        &  4.9  &  4.0  &  3.3           \\
\end{tabular}                                                                           
$^a$ Experimental values from:  \\
J = JANAF\cite{janaf} Tables   \\
H = Huber and Herzberg\cite{Hub79}  \\
C = CODATA\cite{Cod89} Values \\
G = Gurvich \etal\cite{Gurvich} compilation \\
S=standard state          \\
DO = Doncaster, and Walsh, \ijck{13}{503}{1981}  \\
L = Lias \etal\cite{lias} compilation  \\
GO =  W.D. Good, J.L. Lacina, J.P. McCullough, \jpc{65}{2229}{1961}  \\
E = Based on singlet-triplet splitting determined by A.R.W. McKelllar, P.R. Bunker,
T.J. Sears, K.M. Evenson, R.J. Saykally and S.R. Langoff, \JCP{79}{5251}{1983} \\
B = J. Berkowitz, J.P. Greene, H. Cho, and B. Ruscic, \JCP{86}{1235}{1987} \\
$^b$ G2 and G3 values from \cite{g2paper}     \\
$^c$ CBS-Q values from \cite{cbs4}
\end{table}

\newpage

\begin{table}  \linespread{1.0}
\caption{\label{hofdgeetwotwo}
Deviation of heats of formation (kcal/mol) from experiment for selected molecules from the G2-2 test set. }
\squeezetable
\begin{tabular}{ldlcrrrrrr}
         &  \multicolumn{3}{c}{Expt.$^a$} & \multicolumn{5}{c}{Deviation(experiment $-$ theory)}   \\
	 Species  &  \multispan3{\hrulefill}  & \multispan4{\hrulefill}  \\
	          &  \hof\  &  $\pm$(uncert.) & Ref.  & W1   &  W2 &  G2$^b$   &  G3$^b$   &  CBS-Q$^c$  \\
		  \hline
BF$_3$  &  -271.5  &  0.2  & C                        &  1.3    & -0.1  &  -0.1  &  -0.6  &  -1.2               \\
BCl$_3$  &  -96.3  &  0.5  & J                        &  0.4    &  1.0  &  2.0  &  0.0  &  4.0                 \\
CF$_4$  &  -223.0  &  0.2  & G                        &  1.7    &  0.2  &  5.5  &  0.9  &  3.6                   \\
OCS  &  -33.9  &  0.5  & G                            &  0.1    &  0.2  & 1.9  &  2.0  &  3.1                    \\
CS$_2$  &  27.9  &  0.2 & G                           &  0.1    &  0.7  & 2.0  &  3.2  &  5.4                     \\
N$_2$O  &  19.5  &  0.1  & G                          &  -0.8    &-1.2  & -0.7  &  -1.9  &  -0.3                 \\
ClNO  &  12.6  &  0.1 & G                             &  -1.6    & -1.6  &  1.0  &  -0.8  &  1.7                  \\
NF$_3$  &  -31.6  &  0.3 & J                          &  0.1    &       & 3.7  &  0.0  &  2.8                    \\
F$_2$O  &  5.9  &  0.4  & G                           &  -0.8    &-1.0  & 0.5  &  -0.6  &  0.3                    \\
CH$_2$F$_2$  &  -108.1  &  0.2 & G                    &  0.8    &       & 2.7  &  0.3  &  1.3                 \\
CH$_3$CN  &  17.7  &  0.1 & AN                        &  0.3    &       & -0.4  &  -0.1  &  -1.0                 \\
HCOOH  &  -90.5  &  0.1  & G                          &  0.6    &       & 2.0  &  0.1  &  1.1                  \\
C$_2$H$_4$O(oxirane)  &  -12.6  &  0.2 & J            &  -0.2    &      & 1.3  &  0.0  &  0.2                 \\
HCOCOH  &  -50.7  &  0.2 & T                          &  0.7    &       & 2.9  &  0.9  &  1.5                 \\
CH$_3$OCH$_3$  &  -44.0  &  0.1 & T                   &  1.1    &       & 2.0  &  0.4  &  0.5                \\
C$_2$H$_5$Cl  &  -26.8  &  0.2  & T                   &  1.0    &      &  0.8  &  -0.1  &  0.9               \\
CH$_3$COCH$_3$  &  -51.9  &  0.2 & P                  &  1.1    &       &  1.1  &  0.1  &  0.0                \\
HS  &  34.2  &  0.7  & JB                             &  0.5    & 0.6 &  -0.3  &  0.5  & 0.4                     \\
CCH  &  135.9  &  1.2  & G                            &  0.1    &-0.1   &  -2.8  &  -0.4  &  -1.3                 \\
C$_2$H$_3$  &  71.5  &  1.2  & TS                     &  0.9    &       &  -1.2  &  1.0  &  -0.3                  \\
CH$_3$CO  &  -2.9  &  0.7  & TS                       &  0.2    &       &  -0.1  &  -0.4  &  -0.4                \\
CH$_2$OH &  -4.3  &  0.6  & G                         &  0.4    &       &  -0.5  &  -0.4  &  -0.5                \\
CH$_3$CH$_2$O  &  -3.7  &  0.8  & JB                  &  -0.8    &      &  -1.4  &  -1.2  &  -1.9              \\
CH$_3$S  &  29.8  &  0.4  & JB                        &  1.2    &       &  -0.1  &  0.8  &  0.9                   \\
C$_2$H$_5$   &  28.4  &  0.5 & TS                     &  0.4    &       &  -1.5  &  -0.3  &  -1.7                 \\
NO$_2$ &  8.2  &  0.1 & G                             &  -0.8    & -1.2 &  1.0  &  0.1  &  2.6                    \\
\hline
  Mean abs. error &    & 0.4 &                          & 0.7        &&  1.5  &  0.7  &  1.5                               \\
  Max abs. error &     &  1.2&                          & 1.7      && 5.5 &  3.2 &  5.4       \\
\end{tabular}
$^a$ Experimental values from: \\
J = JANAF\cite{janaf} Tables   \\
C = CODATA\cite{Cod89} Values \\
G = Gurvich \etal\cite{Gurvich} compilation \\
T = TRC\cite{trc} compilation \\
TS = Tsang\cite{tsang} compilation \\
AN = X. An, M. Mansson, \jct{15}{287}{1983}   \\
P = Pedley \etal\cite{pedley} compilations \\
JB= Values reported by Berkowitz \etal\cite{Ber94} \\
$^b$ G2 and G3 values from \cite{g2paper}    \\
$^c$ CBS-Q values from \cite{cbs4}
\end{table}
\newpage

\begin{table}
\caption{Computed and observed proton affinities (kcal/mol)\label{PAtable}}

\begin{tabular}{ldddd}
     & Experiment & \multicolumn{3}{c}{Experiment$-$theory}\\
		 & Ref.\cite{LiasPA} & W1 & W2  & G3\\
\hline
NH$_3$ & 204.0 & -0.1 & 0.1 & 0.9\\
H$_2$O & 165.2 & 0.2 & 0.3 & 1.8\\
C$_2$H$_2$ & 153.3 & -0.8 & -0.9 & 0.5\\
SiH$_4$ & 152.9 & -0.3 & -0.2 & 0.6\\
PH$_3$ & 187.6 & 0.4 & 0.5 & 2.3\\
H$_2$S & 168.5 & -0.6 & -0.6 & 1.5\\
HCl & 133.1 & -0.7 & -0.8 & 0.5\\
H$_2$ & 100.9 & -0.4 & -0.4$^a$ & 1.6\\
\hline
Mean abs. error & & 0.44& 0.49& 1.2\\
Max. abs. error & &0.8 & 0.9& 2.3\\
\end{tabular}

(a) Using CCSD(T)/cc-pVQZ anharmonic zero point energies: -0.2 kcal/mol.
(ZPVE(H$_3^+$)=12.56 rather than 12.31 kcal/mol; ZPVE(H$_2$)=6.21 compared
to 6.22 kcal/mol.)

\end{table}

\end{document}